\documentclass[11pt]{article}
\newcommand{\Comment}[1]{{}}
\usepackage{amsfonts,amsthm,amsmath,amssymb,slashed}
\usepackage{bbold}

\usepackage[utf8]{inputenc} 
\usepackage[textwidth = 430 pt, textheight = 630 pt]{geometry}
\usepackage{hyperref,amsthm,amsmath,amssymb,slashed,graphicx,bbold,fancybox,mathtools,color,tikz-cd,caption,subcaption,cite,geometry,marginnote,xcolor,fancybox,verbatim}

\Comment{\usepackage{color}
\definecolor{MyDarkBlue}{rgb}{0.15,0.15,0.45}
\usepackage[linktocpage=true]{hyperref}
\usepackage{makeidx}
\usepackage[numbers,sort&compress]{natbib}
\usepackage{hypernat}}
\usepackage{framed}

\definecolor{MyDarkBlue}{rgb}{0.15,0.15,0.45}
\definecolor{shadecolor}{rgb}{0.85,0.85,0.85}
\definecolor{link}{rgb}{0.15,0.15,0.45}

\hypersetup{
pdftitle={Penrose limits and spin chains in the GJV/CS-SYM duality},
pdfsubject={High Energy Physics},
pdfauthor={Thiago Araujo, Georgios Itsios, Horatiu Nastase and Eoin \'{O} Colg\'{a}in},
pdfkeywords={gauge; susy; strings; fields; cft},
pdfsubject={hep-th},
colorlinks=true,linkcolor=link,citecolor=link,urlcolor=link,linktocpage
}

\Comment{\usepackage{color}
\definecolor{MyDarkBlue}{rgb}{0.15,0.15,0.45}

\usepackage[numbers,sort&compress]{natbib}
\usepackage{hypernat}}
\usepackage{graphicx}

\newcommand\ignore[1]{}
\def\one{{\,\hbox{1\kern-.8mm l}}}

\def\Tr{{\rm Tr\, }}

\def\a{\alpha}\def\b{\beta}
\def\g{\gamma}

\def\m{\mu}
\def\n{\nu}

\def\r{\rho}
\def\s{\sigma}

\def\l{\lambda}

\def\d{\partial}

\def\Z{\mathbb{Z}}

\newcommand{\Cset}{{\,\,{{{^{_{\pmb{\mid}}}}\kern-.45em{\mathrm C}}}}}

\newcommand{\dd}{\mathrm d}

\newcommand{\be}{\begin{equation}}
\newcommand{\bea}{\begin{eqnarray}}

\newcommand{\ee}{\end{equation}}
\newcommand{\eea}{\end{eqnarray}}

\newcommand{\ret}{\nonumber \\}
\newcommand{\nn}{\nonumber}

\newcommand{\bse}{\begin{subequations}}
\newcommand{\ese}{\end{subequations}}

\begin{document}

\renewcommand{\thefootnote}{\fnsymbol{footnote}}

\makeatletter
\@addtoreset{equation}{section}
\makeatother
\renewcommand{\theequation}{\thesection.\arabic{equation}}

\rightline{}
\rightline{}
   \vspace{1.8truecm}

\begin{flushright}
APCTP Pre2017 - 008
\end{flushright}

\vspace{10pt}


\begin{center}
{\LARGE \bf{\sc Penrose limits and spin chains in the GJV/CS-SYM duality}}
\end{center}
 \vspace{1truecm}
\thispagestyle{empty} \centerline{
 {\large \bf {\sc Thiago Araujo${}^{a}$}}\footnote{E-mail address: \Comment{\href{mailto:thiago.araujo@apctp.org}}{\tt thiago.araujo@apctp.org}}, \;
{\large \bf {\sc Georgios Itsios${}^{b}$}}\footnote{E-mail address: {\tt gitsios@gmail.com}}, \;
}
\vspace{.1cm}
\centerline{
   {\large \bf {\sc Horatiu Nastase${}^{b}$}}\footnote{E-mail address: \Comment{\href{mailto:nastase@ift.unesp.br}}{\tt
    nastase@ift.unesp.br}} and
{\large \bf {\sc Eoin \'O Colg\'ain${}^{a}$}}\footnote{E-mail address: {\tt ocolgain.eoin@apctp.org}}
}

\vspace{.8cm}
\centerline{{\it ${}^a$
Asia Pacific Center for Theoretical Physics, POSTECH, Pohang 37673, Korea}}\vspace{.2cm}
\centerline{{\it ${}^b$
Instituto de F\'{i}sica Te\'{o}rica, UNESP-Universidade Estadual Paulista}} \centerline{{\it
R. Dr. Bento T. Ferraz 271, Bl. II, Sao Paulo 01140-070, SP, Brazil}}

\vspace{1.0truecm}

\thispagestyle{empty}

\centerline{\sc Abstract}

\vspace{.4truecm}

\begin{center}
\begin{minipage}[c]{380pt}
{\noindent We examine Penrose limits of the duality proposed by Guarino, Jafferis and Varela between a type IIA massive background
of the type of a warped, squashed $AdS_4\times S^6$, and a 2+1 dimensional IR fixed point of ${\cal N}=8$ super Yang-Mills deformed by Chern-Simons 
terms to ${\cal N}=2$ supersymmetry. One type of Penrose limit for closed strings corresponds to a large charge  closed spin chain, and another, for open strings on giant graviton D-branes, 
corresponds to an open spin chain on sub-determinant operators. For the first limit, we find that like in the ABJM case, there are functions $f_a(\lambda)$
that interpolate between the perturbative and nonperturbative (string) regions for the magnon energy. For the second, we are unable to 
match the gravity result with the expected field theory result, making this model more interesting than ones with more supersymmetry.    
}
\end{minipage}
\end{center}

\today

\vspace{.5cm}

\setcounter{page}{0}
\setcounter{tocdepth}{2}

\newpage

\tableofcontents
\renewcommand{\thefootnote}{\arabic{footnote}}
\setcounter{footnote}{0}

\linespread{1.1}
\parskip 4pt

\newpage

\section{Introduction}


The AdS/CFT correspondence was started in 3+1 dimensions with the duality of ${\cal N}=4$ super Yang-Mills (SYM) and string theory in an $AdS_5\times 
S^5$ background \cite{Maldacena:1997re} (see \cite{Nastase:2015wjb} for an introduction to the correspondence). But many of the results 
obtained there were dependent on the large amount of symmetry, including supersymmetry and conformal invariance, so it was not clear \textit{a priori}
how much of it could be applied to the case of most interest, QCD. Then in 2+1 dimensions the ABJM duality between the ${\cal N}=6$ supersymmetric
$SU(N)\times SU(N)$ superconformal Chern-Simons (CS) gauge theory and strings in an $AdS_4\times S^7 $ background was used as a toy 
model  mostly for 
condensed matter purposes, but again the large amount of symmetry stands in the way of generalizing the lessons observed there to 
physical contexts. 

But an interesting duality was proposed by Guarino, Jafferis and Varela (henceforth GJV) in \cite{Guarino:2015jca} that combines some 
of the best features of both cases, but with less symmetry and more parameters, offering the possibility of getting a little closer to real world 
predictions. On the gravity side it has a 
background solution of the massive type IIA string theory of the type of a warped, squashed $AdS_4\times S^6$, and on the field theory side it 
was proposed that we have an IR fixed point for an ${\cal N}=2$ supersymmetric 2+1 dimensional $SU(N)$ SYM gauge theory obtained from the ${\cal N}=8$ 
one through deforming by a supersymmetric CS terms at level $k$. The CS level is related to the Romans mass $m$ of type IIA by $k=2\pi l_s m$.

Like the ABJM theory, this theory is potentially rich for condensed matter phenomenology (as we said, the ABJM model is the standard toy model
for condensed matter), yet because of its low amount of supersymmetry and extra parameter, can be closer to realistic models. The 
presence of Chern-Simons terms means first of all relevance to anyonic physics (see for instance \cite{Rao:1992aj} for a review). Also, 
since the standard description of the Fractional Quantum Hall Effect (FQHE) involves Chern-Simons terms (see the lectures \cite{Witten:2015aoa}), 
one can hope to embed a holographic description of the FQHE in the GJV background, perhaps modified, like it was done in the case of 
the ABJM model in \cite{Fujita:2009kw, Hikida:2009tp, Bea:2014yda} (description corrected in \cite{Murugan:2013jm}). The ABJM model has been 
also at the center of attempts to describe the quantum critical phase \cite{Myers:2010pk,Mohammed:2012gi,Mohammed:2012rd}, 
and compressible Fermi surfaces \cite{Huijse:2011hp},  but its lack of flexibility (and of parameters) hampered a precise description; perhaps 
now it will have a better chance. 

Another area of interest for theories in 2+1 dimensions with Chern-Simons terms is particle-vortex duality. 
A path integral formulation was started in \cite{Burgess:2000kj},  and developed and embedded in the ABJM model in \cite{Murugan:2014sfa}. 
An $Sl(2,\mathbb{Z})$ action on states \cite{Burgess:2000kj}, including particle-vortex duality, was shown to constrain correlators
of CFTs \cite{Witten:2015aoa}, and similar constraints were found from AdS/CFT in \cite{Herzog:2007ij}. Moreover, 
particle-vortex duality was shown to be part of a web of dualities for 2+1 dimensional theories with Chern-Simons terms, whose basic unit
is a fermionic version of particle-vortex duality \cite{Karch:2016sxi,Seiberg:2016gmd,Murugan:2016zal}, and which can give information 
about condensed matter systems. It is likely that the CS-SYM theory dual to the GJV background can be embedded in a web of dualities also, 
though we have not yet considered this. 

It is well known that the de Wit-Nicolai four dimensional ${\cal N}=8$ $SO(8)$-gauged supergravity can be obtained by a consistent 
truncation of the eleven dimensional supergravity on the seven sphere \cite{deWit:1982bul, deWit:1986oxb, Nicolai:2011cy}. On the 
other hand, it has been shown that this solution belongs to a parametric family of supergravities whose parameter, $\omega\in [0,\pi/8]$, 
is given by a linear combination of the $SO(8)$ electric and magnetic gauge fields. From this perspective, the de Wit-Nicolai solution 
corresponds to the purely electric case \cite{Dall'Agata:2012bb, Dall'Agata:2014ita, deWit:2013ija}.

Naturally, one would like to know if this parametric family of supergravity solutions can be embedded into the string/M-theory framework. 
Unfortunately, it has been recently proved \cite{Lee:2015xga} that we cannot give a general stringy interpretation for this family of 
solutions and that de Wit-Nicolai supergravity solution is an exceptional point in this one-parameter space. Remarkably, there are 
also dyonic supergravity solutions with symmetry group $ISO(7)$, which is closely related to the group $SO(8)$, and with consistent 
embeddings of its purely electric case into the $D=11$ supergravity through consistent truncations on $S^6 \times S^1$ 
\cite{Hull:1988jw, Dall'Agata:2014ita, Guarino:2015qaa}.

In contrast to the $SO(8)$-dyonic solution, it has been shown that the $ISO(7)$-dyonically gauged supergravity solutions can be 
obtained from a massive type IIA solution compactified on a squashed six-sphere, provided that the magnetic coupling constant 
is identified with the Romans mass \cite{Guarino:2015jca, Guarino:2015vca, Varela:2015uca, Pang:2015vna, Pang:2015rwd}. 

Given that the internal manifold is (topologically) a sphere, one can conjecture that the origin of this $ISO(7)$ solution is the 
near-horizon of a stack of $N$ D$2$-branes probing flat space with a Chern-Simons term inducing the Romans mass on the 
brane worldvolume. Using this brane construction, it is conceptually important to replace the transverse flat space by something 
more general, for example,  a Calabi-Yau singularity.

Therefore, taking $N$ D$2$-branes probing a Calabi-Yau threefold singularity in massive type IIA supergravity, the field theory on 
the brane worldvolume is an ${\cal N}=2$ Chern-Simons quiver gauge theory with group $U(N)^G$, where $G$ is the Euler number 
of the resolved manifold, which flows to a field theory dual to a resolution of ${\cal N}=2$ $AdS_4 \times S^6$ in the low energy 
limit \cite{Fluder:2015eoa}.

In a previous paper by two of the authors \cite{Araujo:2016jlx}, the analysis of the GJV/CS-SYM duality was started, with the study 
of various semi-classical observables, such as baryon vertex operators, which are dual to wrapped branes; Wilson loops, the 
anomalous dimensions of operators of high spin coming from large strings, and giant gravitons that are D-branes moving on 
cycles. Furthermore, the analysis of giant magnons was started in the same work.

In this paper, we consider the analysis of spin chains in the duality, based on the model of the closed string spin chain, dual to the Penrose limit,
defined in \cite{Berenstein:2002jq}, and of the open string spin chain, first defined in \cite{Berenstein:2002zw}, applied to the ABJM case, for 
open strings ending on D-branes, in \cite{Cardona:2014ora}. We first find a Penrose limit whose closed string excitations on the pp-wave give a
closed spin chain. We then find Penrose limits for which the open string excitations, for strings ending on D-branes in the pp-wave, give
open spin chains embedded into larger operators. 

The paper is organized as follows. In section 2 we present the GJV/CS-SYM duality: after a review of the gravity solution, 
we discuss in some detail the general CS-SYM field theories, and the particular case of the dual to the gravity solution, with emphasis on the 
IR physics. In section 3 we present the relevant Penrose limits of the GJV geometry. We start with a classification of the useful null geodesics, 
namely ones that move on isometry directions, corresponding to a large charge in the field theory. These are then analyzed 
separately: motion in a direction $\psi$ for closed strings, 
and in possible directions $\sigma,\phi$, and $\sigma+\phi$ for open strings. Then in section 4 we analyze the spin chains
dual to the Penrose limits: we analyze in some detail the one for closed strings, and then we sketch the one for open strings, as 
we could not match properly with the field theory. In section 5 we conclude.  The Appendix contains ${\cal N}=1$ theories and ${\cal N}=2$ supersymmetric 
CS-matter theories in 3 dimensions in superspace.

\section{The duality: $\mathcal{N}=2$ superconformal Chern-Simons theories vs. $AdS_4\times S^6$ background}

The duality proposed in \cite{Guarino:2015jca} is between a well-defined gravitational background, and a CS-SYM gauge theory in 2+1 dimensions, 
which is defined somewhat implicitly, so in this section we will define it better.

We start with a review of the warped, squashed $AdS_4\times S^6$ solution of  \cite{Guarino:2015jca, Fluder:2015eoa}. 
We then consider the relevant $\mathcal{N} = 2$ CS matter theories \cite{Schwarz:2004yj, Gaiotto:2007qi}, and 
specialize them to our case. 

\subsection{Warped, squashed AdS$_4  \times$S$^6$ geometry}

Borrowing the conventions of ref. \cite{Araujo:2016jlx}, the GJV solution  \cite{Guarino:2015jca, Fluder:2015eoa} 
takes the following form in string frame
\bea 
\dd s^2 &=&  e^{{\phi}/{2}+2A} \left( \dd s^2_{AdS_4}+\frac{3}{2}\dd \a^2 + \frac{ 6\sin^2\a}{(3+\cos2\a)} 
\dd s^2_{\mathbb{CP}^2}+ \frac{9\sin^2\a}{(5+\cos2\a)} \eta^2 \right)\label{GJV}, \ret
&\equiv& L^2_{\rm string}\left( \dd s^2_{AdS_4}+\frac{3}{2}\dd \a^2 +\Xi\dd s^2_{\mathbb{CP}^2}+ \Omega \eta^2\right)\cr
e^{\phi} &=& e^{\phi_0} \frac{(5+\cos 2\a)^{3/4}}{(3+\cos 2\a)}, \quad B 
= -\frac{6L^2 e^{\phi_0/2}\sqrt{2}\sin^2\a \cos \a}{(3+\cos2\a)}{\cal J} - \frac{3 L^2 e^{\phi_0/2}}{\sqrt{2}}\sin\a \dd \a\wedge \eta, \ret
\widetilde{F}_0 &=& \frac{1}{ \sqrt{3} L \, e^{5 \phi_0/4}}, \ret 
\widetilde{F}_2 &=& - \frac{\sqrt{6} L}{ e^{3 \phi_0/4}} \left( \frac{4\sin^2\a \cos \a}{(3+\cos 2\a)(5+\cos 2\a)}{\cal J}
+\frac{3(3-\cos 2\a)}{(5+\cos 2\a)^2}\sin \a\; \dd \a \wedge \eta \right), \ret
\widetilde{F}_4 &=& \frac{L^3}{e^{\phi_0/4} } \biggl( 6 \textrm{vol}(AdS_4)-12\sqrt{3}\frac{(7+3\cos2\a)}{(3+\cos2\a)^2}\sin^4\a\; 
\textrm{vol}(\mathbb{CP}^2) \ret
&& \phantom{xxxxxxxxxxxxxx} + 18\sqrt{3}\frac{(9+\cos 2\a)\sin^3\a \cos \a}{(3+\cos 2\a)(5+\cos 2 \a)}{\cal J}\wedge \dd \a \wedge \eta \biggr), 
\eea
where $\a\in [0, \pi]$, $\eta \equiv \dd \psi+\omega$, such that $\dd \omega = 2 {\cal J}$, and we have defined the warp factor
\be
e^{2A}=L^2(3+\cos 2\a)^{1/2} (5+\cos 2\a)^{1/8}.  
\ee
As is common practice in the literature, we denote the Romans mass $m = \widetilde{F}_0$. Explicitly, we write the $AdS_4$ metric as 
\be\label{AdS4} 
\dd s^2_{AdS_4} =-\cosh^2\rho \dd t^2 + \dd \rho^2 + \sinh^2 \rho \dd \Omega^2, 
\ee
the $\mathbb{C P}^2$ metric as
\be
\dd s^2_{\mathbb{CP}^2} = \dd \l^2+\frac{1}{4}\sin^2\l\left\{\dd\theta^2+\sin^2\theta \dd\phi^2+\cos^2\l(\dd\sigma+\cos\theta \dd\phi)^2\right\}, 
\ee
and the one-form connection as 
\be
\label{omega}
\omega = \frac{1}{2} \sin^2 \lambda ( \dd \sigma + \cos \theta \dd \phi). 
\ee
For later convenience, we record the metric on $ \mathbb{CP}^2$ expressed in terms of left-invariant one-forms 
\footnote{We define $\textrm{vol} (\mathbb{CP}^2) = - \frac{1}{2} \mathcal{J} \wedge \mathcal{J}$.}:
\be
\dd s^2_{\mathbb{CP}^2} = \dd \l^2+\frac{1}{4}\sin^2\l\left\{ \tau_1^2 + \tau_2^2 +\cos^2 \l \tau_3^2\right\}. 
\ee
where we have defined
\be
\tau_1 = - \sin \sigma \dd \theta + \cos \sigma \sin \theta \dd \phi, \quad \tau_2 = \cos \sigma \dd \theta 
+ \sin \sigma \sin \theta \dd \phi, \quad \tau_3 = \dd \sigma + \cos \dd \phi. 
\ee
We also take the opportunity to record the field strength $H_3 = \dd B$, 
\be
H_3 = \frac{24 \sqrt{2} L^2 e^{\frac{\phi_0}{2}} \sin^3 \alpha }{[3 + \cos(2 \alpha)]^2} \dd \alpha \wedge \mathcal{J}.
\ee

For constant $\a$ and $\psi$, the internal manifold is topologically $\mathbb{CP}^2$, and for $\l=\pi/2$ and fixed $\s$ we have 
(topologically) a $\mathbb{CP}^1$. The points $\alpha = 0$ and $\alpha = \pi$ correspond to isolated conical singularities. 
The isometries of the metric are $SO(2,3)$ for $AdS_4$ and the $SO(7)$ symmetry of the internal manifold is broken 
down to $SU(3)\times U(1)$ for the $\mathbb{CP}^2$ and the $S^1$ fiber $\psi$ through various warp factors.

The constants in the solution are expressed in terms of the Romans mass $m$ and another parameter $g$ 
\footnote{From the point of view of the four dimensional dyonic supergravity theory, these constants correspond 
to the magnetic and electric couplings, respectively.} by
\be 
L^2\equiv 2^{-5/8}3^{-1}g^{-25/12}m^{1/12}\; \quad \text{and} \quad e^{\phi_0}\equiv 2^{1/4}g^{5/6}m^{-5/6}\; .
\ee
Charge quantization in this background leads to (see \cite{Guarino:2015jca, Fluder:2015eoa, Araujo:2016jlx})
\be
m = \widetilde{F}_0 = \frac{k}{2 \pi \ell_s}, 
\ee
where $k\in \mathbb{Z}$ is an integer that will be associated with the CS level in the field theory dual and 
$\ell_s = \sqrt{\alpha'}$ is the string length, and  allows the relations
\bea
L&=& \frac{\pi^{3/8} \ell_s}{2^{7/48}3^{7/24}}(kN^5)^{1/24} ;\qquad
e^{\phi_0}=  \frac{2^{11/12}\pi^{1/2}}{3^{1/6}}\frac{1}{(k^5 N)^{1/6}} \Rightarrow \cr
L^2_{\rm string}&=&\frac{2^{1/6} \pi}{3^{2/3}}\left(\frac{N}{k}\right)^{1/3}\ell_s^2\sqrt{5+\cos 2\a}\;,\label{quant}
\eea
where $N\in \mathbb{N}$ will be associated with the rank of the gauge group in the field theory dual.

\subsection{CS-SYM field theory action}

The conjecture of \cite{Guarino:2015jca} relates the supergravity solution of \cite{Guarino:2015jca, Fluder:2015eoa} to the IR fixed point of the field theory on a stack of $N$ D2-branes with Romans mass deformation $m$. 

The worldvolume field theory on a stack of $N$ D$2$-branes in flat space is an ${\cal N}=8$ $D=3$ SYM theory with gauge group $SU(N)$, containing 
the gauge field,
$7$ scalars (the transverse directions to the D$2$-brane) and $8$ fermions, all of them in the adjoint representation of the gauge group.

Similarly to the ABJM case, the fixed point for the D2-brane theory on a background with a mass deformation $m$ induces a Chern-Simons term on the 
D$2$-brane worldvolume, 
\be 
S_{CS}=\frac{k}{4\pi}\int \text{Tr}\left(A\wedge \dd A + \frac{2 i}{3}A\wedge A\wedge A \right)\; ,
\ee
where $k\in \mathbb{Z}$ is the Chern-Simons level \cite{Guarino:2015jca, Fluder:2015eoa}, which is related to the Romans mass by $k=2\pi l_s m$.

The Chern-Simons term by itself breaks all the supersymmetries, but by supersymmetrizing it and adding appropriate couplings, 
we can preserve up to ${\cal N}=3$ supersymmetries \cite{Guarino:2015jca, Guarino:2015qaa,Guarino:2015vca, Pang:2015vna, Pang:2015rwd}.

The GJV gravitational solution is an ${\cal N}=2$ background in massive type IIA, so the conjectured field theory dual should have the same amount of supersymmetry. In ${\cal N}=2$ notation, the IR fixed point theory has an $SU(N)$ vector multiplet $V$ and three chiral multiplets $\Phi_i$ for $i=1,2,3$. The theory has a superpotential given by
\be 
{\cal W} = g \Tr\left(\Phi_1[\Phi_2, \Phi_3] \right) = \frac{g}{12}\epsilon_{ijk} f^{abc} \Phi_i^a \Phi_j^b \Phi_k^c\; ,
\ee
with all fields in the adjoint of $SU(N)$. The field content and superpotential are exactly the same as in 4D ${\cal N}=4$  SYM
\cite{terning2009modern, Nastase:2015wjb}.

The theory has an $U(1) \times SU(3)$ symmetry, where the $U(1)=SO(2)$ is the R-symmetry, and $SU(3)$ rotates the complex scalars 
in the chiral multiplets. As we saw, this symmetry is realized in the dual gravitational background as the isometry of the internal space: the 
deformation (squashing) of the six-sphere breaks the original $SO(7)$-isometry down to $SU(3)\times U(1)$, respectively the isometries of  
$\mathbb{CP}^2$ and the $\mathbb{S}^1$ parametrized by the fiber coordinate $\psi$ in the geometry. 

The chiral superfield is expanded as usual into a scalar $\phi$, a fermion $\psi$ and an auxiliary field $F$,
\be
\Phi=\phi+\sqrt{2}\theta\psi+\theta\theta F.
\ee

Note that the dimension of the superpotential coupling is $[g]=1/2$, 
so the superpotential term {\em dominates at low energies (in the IR, close to the conformal 
point)}. Other dimensions are $[D_{\dot\a}]= [ \dd \theta] = 1/2$, $[\theta]=1/2$, whereas $[\Phi]=[\phi]=1/2$.

Superconformal CS theories in 2+1 dimensions have been studied by  Schwarz \cite{Schwarz:2004yj},  Gaiotto and Yin 
\cite{Gaiotto:2007qi}.
The CS matter action in the presence of a superpotential ${\cal W}$ can be written as (see appendix \ref{conventions} for more details) 
\be 
 S = S_{CS} + S_{m} + S_{sp}\; ,
\ee 
and the complete expression is given in (\ref{csm-lag}). The superpotential action is
\begin{align} 
S_{sp} & = - 2\int \dd^3 x \Tr \left( \frac{\d {\cal W}(\phi)}{\d \phi^i} \frac{\d \overline{{\cal W}(\phi)}}{\d \bar{\phi}^i} +  
 \frac{1}{4}\frac{\d^2 {\cal W}(\phi)}{\d \phi^i \d \phi^j}\psi^i \psi^j+\frac{1}{4}\frac{\d^2 \overline{ {\cal W}(\phi)}}{\d \bar{\phi}^i 
 \d \bar{\phi}^j}\bar{\psi}^i \bar{\psi}^j \right)\nn\\
& = -\int \dd^3 x \left( \frac{g^2 \epsilon_{ijk}\epsilon_{ipq} }{8} (f^{abc}\phi_j^b\phi_k^c) (f^{ade}\phi_p^{\dagger d}\phi_q^{\dagger e}) 
+ \frac{g}{4} \epsilon_{ijk}[\psi_i^a (f^{abc}\psi_j^b \phi^c_k)
+ (f^{abc}\phi^{\dagger c}_k \bar{\psi}_j^b ) \bar{\psi}_i^a]
\right)\nn\\
& = -\int \dd^3 x \left( \frac{g^2 \epsilon_{ijk}\epsilon_{ipq} }{4}\Tr([\phi_j,\phi_k] [\phi_p^{\dagger},\phi_q^{\dagger}])
+\frac{g}{2}\epsilon_{ijk}\Tr (\psi_i[\psi_j, \phi_k])+ \frac{g}{2}\epsilon_{ijk}\Tr([\phi^\dagger_i, \bar{\psi}_j]\bar{\psi}_k)\right).
\end{align}
Note that the CS and matter terms in the action are conformal, but the superpotential term is not.

The sextic (conformal) potential term for the scalars in (\ref{csm-lag}),
\bse
\be
H_{\rm int, 1}=-\frac{16\pi^2}{k^2}\Tr\left(\phi^{i\dagger} T^a \phi^i\right) \Tr\left(\phi^{j\dagger} T^b \phi^j\right) \Tr\left(\phi^{k\dagger} T^a T^b \phi^k\right)\; ,
\ee
can be rewritten as
\be
H_{\rm int, 1}=\frac{4\pi^2}{k^2}\Tr\left([[\phi^{i\dagger},\phi^i],\phi^{k\dagger}][[\phi^{j\dagger},\phi^j],\phi^k]\right)\; .\label{intpot}
\ee
\ese
The other term in the scalar potential is the non-conformal one that comes from the superpotential, which, since 
$\epsilon^{ijk}\epsilon_{ipq}=2\delta_{pq}^{jk}$, is 
\be
\frac{g^2}{2}\Tr\left([\phi_i,\phi_j][\phi^{i\dagger},\phi^{j\dagger}]\right).
\ee

\noindent
{\bf Comments on dimensions and the IR fixed point}

\noindent
In order to understand the IR fixed point of the above theory, we must understand the dimensions of various quantities relevant in the IR, 
and what is the interaction term relevant in the IR.

Gaiotto and Yin \cite{Gaiotto:2007qi} consider the case of a system of D2-branes and D6-branes in massive type IIA theory, 
with a superfield $\Phi_1$ corresponding to the D2-brane coordinates
transverse to the D6-brane (overall transverse), and $\Phi_2,\Phi_3$ to the D2-brane coordinates parallel to the D6-brane (relative transverse), 
whereas $Q,\tilde Q$ are the ``bifundamental", or D2-D6, coordinates. Then they
consider the superpotential {\em in the IR}
\be
{\cal W}=\Tr[\Phi_1[\Phi_2,\Phi_3]]+\tilde Q \Phi_1 Q\;,
\ee
where, due to quantum corrections, in the IR we have $\Phi_1$ of dimension 1 (whereas the fields $\Phi_2,\Phi_3,Q,\tilde Q$ have the classical
dimension 1/2), so that ${\cal W}$ is a marginal operator, i.e. it has the classical dimension of 2. 

But this is only possible because in the IR $\Phi_1$ is 
auxiliary, i.e. it has lost its kinetic term $\int \dd^4\theta \bar\Phi_1\Phi_1$, which would have meant (since $\theta$ has always dimension 1/2) the 
classical dimension 1/2. Then in fact we can introduce a further auxiliary term $\epsilon\Tr\Phi_1^2/2$, which means that by eliminating it we 
obtain the usual quartic potential for $\Phi_2,\Phi_3,Q,\tilde Q$, 
\be
{\cal W}=\frac{1}{2\epsilon}\Tr[([\Phi_3,\Phi_3]+\tilde Q Q)^2].
\ee

But that was only possible since we have singled out $\Phi_1$, as being the superfield for the coordinates transverse to the D6-branes (overall transverse), 
and to the fact that in the IR,
quantum corrections dominate and kill the kinetic term for $\Phi_1$. 
In their absence, this should not be possible. Then $\Phi_1,\Phi_2,\Phi_3$ should appear symmetrically in the action, and this is the case that 
we have now. 

In fact, \cite{Guarino:2015jca} argue that $\Phi_1,\Phi_2,\Phi_3$ have R-charge $q_R=2/3$, which would mean that ${\cal W}$ has R-charge 2. 
As an operator, ${\cal W}$ can stay chiral if $\Delta=q_R$, so that would mean that the dimensions of $\Phi_1,\Phi_2,\Phi_3$ are also 2/3, for a total 
dimension of 2. However, then, while $\int \dd^2\theta {\cal W}$ has still dimension 3, as needed ($\int \dd^2 \theta $ always has the classical dimension), this
would only generate the 
\be
\frac{\d {\cal W}}{\d \phi^u}F^i
\ee
term in the action, with the understanding that $F^i$ has the dimension $[\Phi]-[\theta\bar\theta]=[\Phi]+1=5/3$. But that is only possible, again, 
if there is no kinetic term $\int \dd^4\theta \bar\Phi\Phi$ in the action, i.e. if the quantum corrections have renormalized it away, by multiplying it 
with a factor $\mu^\delta$, where $\mu$ is the renormalization scale and $\delta$ the anomalous dimension of the kinetic operator. 
Besides losing the dynamics of $\Phi_i$, this would mean that now there is no $F_i^2$ term anymore, so eliminating $F_i$ we now obtain instead  
\be
\frac{\d {\cal W}}{\d \phi_i}=0\Rightarrow [\phi_i,\phi_j]=0.
\ee

But that is the same condition as would be obtained by considering instead a superpotential with coefficient $g$ of dimension 1/2, 
understood in the quantum theory as having dimension coming from a  $\mu^\delta$ factor, which therefore 
would dominate at low energies. Then the potential coming from it must be put to zero at low energies, again obtaining 
\be
[\phi_i,\phi_j]=0.
\ee
The relation between the two pictures described above is a rescaling of the $\Phi_i$'s by  $g^{1/3}$, 
which would imply that after it, the kinetic term has a coefficient with dimension. 
Either way, the result is the same, namely $\phi_i$'s should commute in order to avoid having an infinite potential term in the IR.

But that still leaves us with the conformal term in the potential, which survives the IR limit unchanged. This is given in (\ref{intpot}).

Then in the IR (at low energies), for the picture with mass dimension $[g^2]=1$,  
there will be no conformal point unless the commutator of $\phi_i$'s vanishes,
which means that we will be restricted to live on the space of solutions with 
\be
[\phi_i,\phi_j]=0,\;\;\; \forall ~ i\neq j.
\ee
Note that this still leaves the possibility that $[\phi_i,\bar\phi_i]\neq 0$. Indeed, this is needed in order to have the conformal term in the potential be 
nonzero.\footnote{Indeed, note that for instance 
\be
\left[\begin{pmatrix} 1 & a\\0&1\end{pmatrix},\begin{pmatrix}1&0\\a^*&1\end{pmatrix}\right]=|a|^2\begin{pmatrix} 1 & 0\\0&-1\end{pmatrix}\;,
\ee
as an example of such a case for $N=2$ (in the $SU(2)$ gauge group case). 
Moreover, it is possible to have also $[\phi_i,\phi_j]=0$, yet $[\phi_i,\bar\phi_j]\neq 0$, for instance 
\be
\left[\begin{pmatrix} 1 & a \\ 0 &1\end{pmatrix},\begin{pmatrix} 1& b\\0& 1\end{pmatrix}\right]=0\;,
\ee
yet
\be
\left[\begin{pmatrix} 1 & a \\ 0 &1\end{pmatrix},\begin{pmatrix} 1& 0\\b^*& 1\end{pmatrix}\right]=ab^*\sigma_3.
\ee}

\section{Penrose limits of the GJV background}
In this section we will study various Penrose limits\cite{Berenstein:2002jq, Gaiotto:2008cg} \footnote{See also 
\cite{Blau:2001ne, Blau:2002dy, Blau:2002mw, Gauntlett:2002cs, Corrado:2002wi, Alishahiha:2002nf, Sugiyama:2002tf, Sadri:2003ib, Park:2012it, Nastase:2015wjb} 
for a non-exhaustive list of references.} of the GJV background from the last section. For any Penrose limit, 
near a null geodesic moving in any direction $x$ in the background, we can consider in principle closed strings or open strings and quantize them (find the 
worldsheet Hamiltonian). 
Closed strings would be dual to a spin chain that selects the scalar $Z$ dual to the direction $x$ as special.

Alternatively, we can consider giant gravitons, i.e. D-branes wrapping some cycle and moving at the speed of light, and the Penrose limit near a null 
geodesic moving in a direction $y$
{\em along} the giant gravitons. Then consider open strings ending on the D-brane, in the Penrose limit, i.e. open string states on the pp-wave, 
corresponding to open strings moving on the D-brane in this direction $y$.

Either way, the starting point for all these exercises is finding the Penrose limit near a null geodesic moving in some direction in the 
background. In this section, we turn our attention to this task and study various Penrose limits of the geometry (\ref{GJV}). 
Case by case, we find it convenient to shift the $\omega$ term by a constant piece that does not change $\mathcal{J}$, so it does not affect the solution.  

\subsection{Useful null geodesics for Penrose limit}

The only thing we need strictly speaking in order to define a Penrose limit is a null geodesic. We will however also consider 
the concept of a ``useful limit", which will mean for us a Penrose limit in an isometry direction.  This should correspond in the 
dual field theory to a spin chain that singles out a large charge $J$ for the corresponding field theory symmetry.

The equation of motion for a null geodesic parametrized by $\lambda$, moving in 10D spacetime with coordinates $x^i$, is 
\be
0=\frac{\dd^2x^i}{\dd \lambda^2}+{\Gamma^i}_{jk}\frac{\dd x^j}{\dd \lambda}\frac{\dd x^k}{\dd \lambda}=
\frac{\dd^2 x^i}{\dd \lambda^2}+\frac{1}{2}g^{il}(\d_k g_{lj}+\d_j g_{lk}-\d_l g_{jk})\frac{\dd x^j}{\dd \lambda}\frac{\dd x^k}{\dd \lambda};\; \forall ~ i=0,1,...,9.
\ee
If we have motion (velocity) in the direction $x^\lambda$, that is $\dd x^i/\dd \lambda=\delta^i_\lambda$, 
we need have no acceleration in the other directions, so 
\be
{\Gamma^i}_{\lambda\lambda}=0\Rightarrow 2 g^{il}\d_\lambda g_{l\lambda}-g^{il}\d_l g_{\lambda\lambda}=2g^{il}\d_\lambda g_{l\lambda}
-\d^i g_{\lambda\lambda}=0.
\ee
Note that we also have $\dd t/\dd \lambda=c$ (constant), but we will deal with {\em static} metrics, $\d_t g_{ij}=0$, and also diagonal metrics with 
$g_{0i}=g^{0i}=0$. Thus, it is easy to see that the geodesic equation for $i=t$ is satisfied, and moreover $j,k=t$ does not contribute to the 
equations for $i\neq t$.

Moreover, if we consider motion in an {\em isometry} direction, i.e. a direction for which $\d_\lambda g_{\mu\nu}=0$, then the condition becomes simply
\be
g^{il}\d_l g_{\lambda\lambda}=\d^i g_{\lambda\lambda}=0.\label{geodcond}
\ee

The three internal isometry directions of the GJV metric are $\sigma,\psi,\phi$, since as we can see, $\d_\sigma g_{\mu\nu}=\d_\phi g_{\mu\nu}=\d_\psi g_{\mu\nu}=0$.

Thus in our case, the metric is a matrix in the $(\sigma,\phi,\psi)$ (isometries) space, and is diagonal in the $(\a,\lambda,\theta)$ 
(non-isometries) space. Specifically, we have
\bea
\frac{g_{\sigma\sigma}}{L^2_{\rm string}(\a)}&=&\frac{\Xi}{4}\sin^2\lambda\cos^2\lambda+\Omega\left(\frac{\sin^2\lambda}{2}-\frac{1}{4}\right)^2=
fct.(\a,\lambda)\cr
\frac{g_{\sigma\phi}}{L^2_{\rm string}(\a)}&=&\left(\frac{\Xi}{4}\sin^2\lambda\cos^2\lambda+\frac{\Omega}{4}\sin^4\lambda\right)\cos\theta
=fct.(\a,\lambda,\theta)\cr
\frac{g_{\phi\phi}}{L^2_{\rm string}(\a)}&=&\frac{\Xi}{4} \sin^2\lambda (\sin^2\theta+\cos^2\lambda\cos^2\theta)
+\frac{\Omega}{4}\sin^4\lambda \cos^2\theta=fct.(\a,\lambda,\theta)\cr
\frac{g_{\psi\phi}}{L^2_{\rm string}(\a)}&=&\frac{\Omega}{2}\sin^2\lambda \cos\theta=fct.(\a,\lambda,\theta)\cr
\frac{g_{\psi\sigma}}{L^2_{\rm string}(\a)}&=&\frac{\Omega}{2}\sin^2\lambda=fct.(\a,\lambda)\cr
\frac{g_{\psi\psi}}{L^2_{\rm string}(\a)}&=&\Omega=fct.(\a).
\eea
Note, in displaying the above metric, we have allowed for a shift $\omega\rightarrow \omega-\dd\sigma/4$, which doesn't change the solution. 

While we can consider in principle the Penrose limit around {\em any } null geodesic, it is more useful to consider the motion around null geodesics
in isometry directions, since that guarantees, as we said, that in the dual field theory we have a spin chain with some large charge $J$ associated
with a symmetry direction matching the isometry of the geodesic. As we emphasized, we can consider closed strings for Penrose limits in any of the ($\sigma,\phi,\psi$) directions, and they would correspond to 
spin chains with some large charge in the field theory. 

But in particular, we will be interested in the Penrose limit for the motion in $\psi$, since as we said, this is the $U(1)$ isometry corresponding in the 
field theory to the $U(1)$ R-symmetry. Thus, we will be considering
closed strings, and giant gravitons, i.e. D4-branes wrapping the $\mathbb{CP}^2$ and moving 
at the speed of light, both in the $\psi$ direction. These objects have to be situated at a point in $AdS_4$ (i.e. we consider the null geodesic 
fixed at a point in $AdS_4$), usually taken to be the center, $\rho=0$. The rest 
of the conditions on the position of the null geodesic need to be defined by the need to get a nontrivial pp-wave (corresponding to a nontrivial 
spin chain in field theory) and by the solutions to the geodesic conditions (\ref{geodcond}). 

Given the above considerations, the motion of the open strings attached to the giant graviton is described by the Penrose limit of null geodesics around 
{\em another isometry direction}, one that can be considered parallel to the D4-branes, i.e. along the $\mathbb{CP}^2$. This means 
either $\sigma$ or $\phi$, or even $\sigma+\phi$, or some other combination of them. 
For both motion in $\sigma$ and $\phi$, it is natural to consider an expansion around 
$\theta=\pi/2$, since in the $\phi$ case we want the coefficient $\sin^2\theta$ of the free $\dd \phi^2$ (the one not mixing with $\dd \sigma$) to be nonzero, and 
more specifically extremum (maximal), and in the $\sigma$ case, we want $\sigma$ not to mix with $\phi$, so $\cos\theta=0$. On the other hand, for motion in 
$\sigma+\phi$, we want to have this combination in the metric, so we need $\cos\theta=1$, i.e. $\theta=0$. 

{\bf Case 1: motion in $\psi$}

We remind the reader that here we consider the shifted $\omega$, i. e. $\omega \rightarrow \omega - 1/4 \dd \sigma$. Now $g_{\psi\psi}$ is, as we saw, only a function of $\a$, so the geodesic conditions reduce to 
\be
g^{\a\a}\d_\a g_{\psi\psi}={\frac{2}{3} }\frac{1}{L^2_{\rm string}}\d_\a ( L^2_{\rm string}\Omega )=0\;,
\ee
which implies,  
\be
\Omega \d_\a\ln L^2_{\rm string }+\d_\a\Omega={ \frac{9 ( 22 \sin 2 \alpha + \sin 4 \alpha)}{4(5 + \cos 2 \alpha)^2} }=0\,.
\ee
The solutions to this equation are $\a=0,\pi/2,\pi$, but in order to have a nontrivial Penrose limit we need to have a nonzero metric $g_{\psi\psi}$, 
which means $\sin^2\a=1$. 

That leaves only $\a=\pi/2$ as a possibility. As usual, we choose also $\rho=0$, meaning that the geodesic is fixed at the center of $AdS_4$, 
and $\lambda=0$, though any $\lambda_0$ would do.

{\bf Case 2: motion in $\sigma$} 

The geodesic conditions, 
\be
g^{\phi\phi}\d_\phi g_{\sigma\sigma} = g^{\psi\psi}\d_\psi g_{\sigma\sigma}=0 \,, 
\ee
are automatically satisfied since $\psi$ and $\phi$ are isometric directions, and 
\be
g^{\theta\theta}\d_\theta g_{\sigma\sigma}=0
\ee
(as well as the similar ones in the AdS directions) are satisfied since $\d_\theta g_{\sigma\sigma}=0$. 

We are left to satisfy the conditions:
\bea
g^{\a\a}\d_\a g_{\sigma\sigma}
&=& {\frac{1}{6} }\left[ - \left(\sin^2\lambda\cos^2\lambda\frac{6\sin^2\a}{3+\cos 2\a}+\left(\sin^2\lambda-\frac{1}{2}\right)^2\frac{9\sin^2\a}{5+\cos 2\a}\right)
\frac{\sin 2\a}{5+\cos 2\a}\right.\cr
&&\left.+\sin^2\lambda\cos^2\lambda\left(\frac{12\sin^2\a\sin 2\a}{(3+\cos 2\a)^2}+\frac{6\sin 2\a}{3+\cos 2\a}\right)\right.\cr
&&\left.+\left(\sin^2\lambda -\frac{1}{2}\right)^2\left(\frac{18 \sin^2\a\sin 2\a}{(5+\cos 2\a)^2}+\frac{9\sin 2\a}{5+\cos 2\a}\right)\right]=0\;,
\eea
which has the solution $\sin 2\a=0$, so $\a=0$ or $\a=\pi/2$, and the condition 
\bea
g^{\lambda\lambda}\d_\lambda g_{\sigma\sigma}
=  \frac{\sin^2 \alpha \sin 4 \lambda}{8 ( 5 + \cos 2 \alpha)}   = 0 \, , 
\eea
which has the solution $\alpha = 0, \pi$ or $\sin 4 \lambda = 0$, so that $\lambda = \pi /4$. 

But we need $\sin^2\lambda \cos^2\lambda\neq 0$, in order to have a nontrivial metric for $\sigma$, along which we move. This 
selects $\lambda=\pi/4$ as the unique solution for the second equation above. We also need $\sin^2\a\neq 0$ for the same reason, which selects $\a=\pi/2$ as the unique solution for the first equation.

All in all, we see that the unique solution for motion in $\sigma$ is $\lambda=\pi/4$, $\a=\pi/2$, $\theta=\pi/2$, $\rho=0$. Note also that 
having $\theta=\pi/2$, from $\dd \theta^2+\sin^2\theta \dd \phi^2$, it follows that we need also to fix $\phi=0$ for the geodesic (actually, any $\phi=\phi_0$ 
would do, since it is an isometry, but it makes no difference). Also we need $\psi=\psi_0$, which again we can choose $\psi=0$. 

{\bf Case 3: Motion in $\phi$} \\
In this case, we consider the original unshifted $\omega$ (\ref{omega}).  The geodesic conditions 
\be
g^{\sigma\sigma}\d_\sigma g_{\phi\phi}= 
g^{\psi\psi}\d_\psi g_{\phi\phi}=0, 
\ee
are automatically satisfied since $\psi$ and $\sigma$ are isometric directions, and 
\be
g^{ij}\d_j g_{\phi\phi}=0\, ,
\ee
where $i$ are the AdS directions, are satisfied since $\d_i g_{\phi\phi}=0$.  

The condition
\be
g^{\theta\theta}\d_\theta g_{\phi\phi}=0, 
\ee
implies 
\be
0 =  \sin^2 \alpha \sin 2 \theta \sin^2 \lambda \; , 
\ee
which has the solutions $\alpha = 0$, $\theta = 0, \pi/2$ and $\lambda = 0$. Observe that shifting $\omega$ does not increase the possible solutions, so we have not considered it.  

The condition 
\be
g^{\lambda\lambda}\d_\lambda  g_{\phi\phi}=0
\ee
implies
\bea
\sin 2\lambda \left(\sin^2\theta+\cos^2\theta\cos 2\lambda +2\frac{\Omega}{\Xi}\cos^2\theta\sin^2\lambda\right)=0\;,
\eea
which has the only solution $\sin 2\lambda=0$, i.e. $\lambda =0$ or $\pi/2$. Again, we remark that shifting $\omega$ does not help increase the 
number of solutions. 

The condition 
\be
g^{\a\a}\d_\a g_{\phi\phi}=0
\ee
becomes 
\bea
&& - \frac{1}{4}[\Xi\sin^2\lambda(\sin^2\theta +\cos^2\lambda\cos^2\theta)+\Omega \sin^4\lambda\cos^2\theta]\frac{\sin 2\a}{5+\cos 2\a}\cr
&&+ {6} \sin^2\lambda(\sin^2\theta +\cos^2\lambda\cos^2\theta) { \frac{\sin 2\a}{(3+\cos 2\a)^2} } \cr
&&+ {\frac{27}{2}}\sin^4\lambda \cos^2\theta  { \frac{\sin 2\a}{(5+\cos 2\a)^2} }=0\;,
\eea
which has as solutions $\sin 2\a=0$ or $\sin \lambda=0$, i.e. $\a=0$ or $\pi/2$ or $\lambda=0$. 

We now see that $\lambda=\pi/2$ solves the second equation. Indeed, since we need $\sin^2\lambda$ (from the coefficient of the metric in 
$\phi$ direction) to be nonzero, $\lambda=\pi/2$ is the unique valid solution. We also need $\sin^2\a\neq 0$ for the same reason, 
which means that the unique valid solution to the last equation is $\a=\pi/2$. 

All in all, in this case, we need to expand around the geodesic with $\lambda=\pi/2$, $\a=\pi/2$, $\theta=\pi/2$. Also as usual, $\rho=0$ 
(the center of AdS), is chosen by convention (we can always change coordinates in order to put the center where we want).

{\bf Case 4: Motion in $\phi+\sigma$}

Note that in this case it is natural to take $\theta=0$. However, we will obtain this condition from the null geodesic conditions. We first define
\be
\sigma' =  \frac{\sigma+\phi}{\sqrt{2}};\;\;\; \phi'=\frac{\sigma-\phi}{\sqrt{2}}\;,
\ee
so that, also replacing now $\omega\rightarrow \omega - \dd \sigma/4- \dd \phi/4$, we have
\bea
\dd s^2_{\mathbb{CP}^2}&=&\dd \lambda^2 + {\frac{\sin^2 \lambda}{4} \dd \theta^2}+\frac{\sin^2\lambda}{8}\biggl( [\sin^2\theta+\cos^2\lambda (1+\cos\theta)^2]  \dd \sigma'^2 \cr && 
+[\sin^2\theta +\cos^2\lambda(1-\cos\theta)^2] \dd \phi'^2-2 \sin^2\theta\sin^2\lambda \dd \sigma' \dd \phi' \biggr) \;,  \cr
\eta&=&\dd \psi+\frac{(\sin^2\lambda-1/2)}{2}\frac{\dd \sigma' {+} \dd \phi'}{\sqrt{2}}+\frac{(\sin^2\lambda \cos\theta-1/2)}{2}\frac{\dd \sigma'-\dd \phi'}{\sqrt{2}}.
\eea

The null geodesic conditions are then
\be
g^{il}\d_l g_{\sigma' \sigma'}=0\;,
\ee
where
\be
\frac{g_{\sigma'\sigma'}}{L^2_{\rm string}(\a)}=\frac{\Xi}{8}\sin^2\lambda [\sin^2\theta+\cos^2\lambda(1+\cos\theta)^2]
+\frac{\Omega}{8}(\sin^2\lambda(1+\cos\theta)-1)^2.
\ee

The conditions
\be
g^{\phi'\phi'}\d_{\phi'}g_{\sigma'\sigma'}= 
g^{\psi\psi}\d_\psi g_{\sigma'\sigma'}=0
\ee
are satisfied since $\phi'$ and $\psi$ are isometries, and 
\be
g^{ii}\d_i g_{\sigma'\sigma'}=0
\ee
since $g_{\sigma'\sigma'}$ is independent on the AdS coordinates $i$. Then 
\bea
 g^{\theta\theta}\d_\theta g_{\sigma'\sigma'}
=  - \frac{\sin^2 \alpha \sin \theta [ \cos^2 (\theta/2) \cos 2 \lambda + \sin^2 (\theta/2)]}{5 + \cos 2 \alpha} , 
\eea
has the unique solution $\sin\theta=0$, i.e. $\theta=0$. Also,
\bea
g^{\lambda\lambda}\d_\lambda g_{\sigma'\sigma'}
=  \frac{\cos^2 ( \theta/2) \sin^2 \alpha \sin 2 \lambda [ \cos 2 \lambda + \tan^2 ( \theta/2) ] }{2 ( 5 + \cos 2 \alpha)} , 
\eea
when evaluated at $\theta=0$, becomes
\be
 \frac{ \sin^2 \alpha \sin 4 \lambda}{4 (5 + \cos 2 \alpha)} = 0, 
\ee
with solutions $\sin 4\lambda=0$, i.e. $\lambda =0,\pi/4,\pi/2,3\pi/4$  and $\alpha = 0$. 

But since we want $\sin^2\lambda\cos^2\lambda\neq 0$ for a nontrivial Penrose limit, we must choose $\lambda=\pi/4$.
Finally, the remaining geodesic condition, 
\bea
g^{\a\a}\d_\a g_{\sigma'\sigma'}
&=& - \frac{1}{8}[\Xi(\sin^2\lambda\sin^2\theta+\cos^2\lambda (1+\cos\theta)^2+\Omega(\sin^2\lambda(1+\cos\theta)-1)^2]\frac{\sin 2\a}{5+\cos 2\a}\cr
&&+{\frac{1}{3}} [\sin^2\lambda\sin^2\theta+\cos^2\lambda(1+\cos\theta)^2] {\frac{ \sin 2\a}{(3+\cos 2\a)^2} }\cr
&&+ {\frac{27}{4}} [\sin^2\lambda(1+\cos\theta)-1]^2 { \frac{\sin 2\a}{(5+\cos 2\a)^2}=0 }\;,
\eea
has the solution $\sin 2\a=0$, or $\a=0, \pi/2$. Since we want $\sin^2\a\neq 0$, we must use $\a=\pi/2$. 

In conclusion, the geodesic is $\theta=0$, $\lambda=\pi/4$, $\a=\pi/2$, as well as $\rho=0$ and $\psi=0$ as valid choices (we could use 
$\psi_0$).

\subsection{Useful pp-wave for closed strings: motion in $\psi$}
In this subsection we consider the Penrose limit of the GJV solution near a null geodesic moving on the $U(1)$ isometry coordinate $\psi$, which is dual to R-symmetry. As highlighted in the previous section (case 1) we should expand around $\a=\pi/2$, $\l=0$ and $\r=0$. Before proceeding, it is worth noting that there are no pp-wave solutions to a massive theory, so in the process of zooming in on the null geodesic, the final solution we encounter is a solution to massless IIA supergravity. Therefore, ensuring that the equations of motion of the theory are satisfied provides an important consistency check that we have performed the limit correctly.

To perform the Penrose limit, we first rescale as usual the coordinates in the vicinity of the null geodesic, consistent with $\a=\pi/2,\lambda=0,\rho=0$, for the 
$R \equiv  e^{\frac{\phi_0}{4}} L \rightarrow \infty$ limit (the near-geodesic limit) \footnote{Observe 
that the limit $R\to \infty$, corresponds to the case where $(m,g) \to 0$, with 
$e^{\phi_0}\sim \left(\frac{g}{m}\right)^{5/6}$ is fixed and $L^2\sim \left(\frac{m}{g^{25}}\right)^{1/12}\to \infty$.}
\begin{align}
t & =  \frac{\tilde{t}}{\sqrt{2}}\ ,\quad \psi = \frac{\sqrt{2}}{3} \tilde{\psi}\, \qquad \tilde{t} = x^+ + \frac{x^-}{R^2}, \quad \tilde{\psi} = x^+ - \frac{x^-}{R^2}, \nn \\
\a & = \frac{\pi}{2} + \frac{u}{\sqrt{3} R}\ , \qquad \rho=\frac{\tilde{\rho}}{\sqrt{2} R}\ , \qquad \l=\frac{x}{\sqrt{6} R}\; .
\end{align}
Taking the limit $R\to \infty$, the metric reduces to 
\bse
\be
\begin{split}
\dd s^2_{pp}  = & -4 \dd \tilde{x}^+ \dd \tilde{x}^- +   \dd u^2 + \dd \tilde{\rho}^2 + \tilde{\rho}^2 \dd \Omega_2^2  +  \dd x^2 + \frac{x^2}{4}\sum_{j=1}^3 \tau_j^2\\ 
&-\left( \frac{u^2}{2} +\frac{ \tilde{\rho}^2}{2}\right) (\dd \tilde{x}^+)^2 + \frac{x^2}{2\sqrt{2}}\tau_3 \dd \tilde{x}^+\; .
\end{split}
\ee

Changing from spherical $(\tilde{\r},\vartheta, \varphi)$ to Cartesian coordinates $(y_1,y_2,y_3)$, and using the Hopf map coordinates $(z_1,z_2)= \left(x\sin \frac{\theta}{2}e^{\frac{i}{2}(\s-\phi)}, x\cos \frac{\theta}{2}e^{\frac{i}{2} (\s+\phi)}\right)$ for $\mathbb{C}^2$, we have
\be
\begin{split}
\dd s^2_{pp}  = & -4 \dd \tilde{x}^+ \dd \tilde{x}^- +   \dd u^2 + \sum_{i=1}^3 (\dd y_i )^2  + \sum_{j=1}^2 \dd z_j \dd \bar{z}_j\\ 
&-\left( \frac{u^2}{2} + \frac{1}{2}\sum_{i=1}^3 y_i^2\right) (\dd \tilde{x}^+)^2 - \frac{i}{2\sqrt{2}}\sum_{j=1}^2\left(\bar{z}_j \dd z_j - z_j \dd \bar{z}_j \right) \dd \tilde{x}^+\; .
\end{split}
\ee
\ese
In order to write this metric in the standard Brinkmann pp-wave form \cite{Brinkmann:1925fr}, so that it is quadratic in transverse coordinates, 
\be
\label{Brinkmann}
\dd s^2_{pp}  = -4 \dd x^+ \dd x^- +  \sum_{i=1}^8 \dd X_i^2   + A_{ij}(x^+) X_i X_j (\dd x^+)^2\; ,
\ee
we consider the scaling 
\be
\tilde{x}^+= \sqrt{2}x^+, \quad  \tilde{x}^-= \frac{x^-}{\sqrt{2}}, \quad  z_j=e^{-\frac{i}{2} x^+}w_j\; , \quad \bar{z}_j=e^{\frac{i}{2} x^+}\bar{w}_j. \label{zredef}
\ee
Finally, we get
\be
\begin{split}
\dd s^2_{pp} & =  -4 \dd x^+ \dd x^- +   \dd u^2  + \sum_{i=1}^3 (\dd y_i )^2  + \sum_{j=1}^2 \dd w_j \dd \bar{w}_j \\
&-\left( u^2 + \sum_{i=1}^3 y_i^2 +\frac{1}{4}\sum_{j=1}^2 |w_j|^2\right) (\dd x^+)^2
\; .
\end{split}
\ee
We can also take the same limit on the remaining fields, with the result, 
\bse
\begin{align}
e^{\phi} &= \sqrt{2} e^{\phi_0}, \quad H_3   =  0, \nn \\
\widetilde{F}_0 & =0, \quad 
\widetilde{F}_2  = -\frac{e^{-\phi_0}}{\sqrt{2} }\dd u\wedge \dd x^+, \quad 
\widetilde{F}_4  = \frac{3 e^{-\phi_0} }{\sqrt{2}}\dd x^+ \wedge \dd y_1\wedge \dd y_2\wedge \dd y_3\; .
\end{align}
\ese
To confirm that there is no mistake, it is prudent to check the geometry is indeed a solution to massless IIA supergravity. For our purposes, we will confine our attention to the Einstein equation, 
\be
\label{Einstein}
R_{\mu \nu} + 2 \nabla_{\mu} \nabla_{\nu} \phi - \frac{1}{4} H_{3 \,\mu \nu}^2  = e^{2 \phi} \left[ \frac{1}{2} (\tilde{F}_2)^2_{\mu \nu}  + \frac{1}{12} (\tilde{F}_4^2)_{\mu \nu} - \frac{1}{4} g_{\mu \nu} \left( \frac{1}{2} \tilde{F}_2^2 + \frac{1}{24} \tilde{F}_4^2 \right) \right],  
\ee
where further details of notation can be found in \cite{Itsios:2012dc}. The immediate advantage of Brinkmann coordinates is that the only nonzero component of the Ricci tensor is $R_{++} =  - \frac{1}{2} \nabla^2 g_{++}$, where we have introduced the Laplacian on the eight-dimensional space transverse to the null-coordinates $(x^+, x^-)$. A quick calculation reveals that $R_{++}=5$ and this is the only nonzero term on the LHS of equation (\ref{Einstein}). Evaluating the RHS, one confirms the same result, so that the Einstein equation is satisfied.

However, even without checking the Einstein equation, we know the solution is correct. In particular, it can be checked that the above pp-wave solution is the same as the Penrose limit of the $AdS_4\times \mathbb{CP}^3$ spacetime \cite{Sugiyama:2002tf,Hyun:2002wu}, which allows us to import the following  
analysis from the literature \cite{Gaiotto:2008cg, Cardona:2014ora}.

The metric, which is warped product of $AdS_4$ with a squashed $S^6$, has isometry  $SO(2,3)\times SU(3)\times U(1)$, with the $U(1)$ R-symmetry. 
In the Penrose/large $R$-charge limit, the $U(1)$ combines with 
the $SO(2)$ (scaling) part of $SO(2,3)$, becoming the $U(1)_{\Delta+R}\equiv U(1)_\pm$ of the field theory.
The Penrose limit of the gravitational background rearranges and breaks the isometry into 
\be
U(1)_\pm \times SO(3)_r \times U(1)_u \times SO(3)\rightarrow U(1)_R\times SU(2)_r\times U(1)_u\times SU(2)_L . 
\ee
Here $SO(3)_r=SU(2)_r$ rotates the coordinates $y_1,y_2,y_3$, $U(1)_u$ gives translations along $u$ and $SU(2)_L$ acts on the 
complex coordinates $w_1,w_2$. Note that these would be four real coordinates, acted upon by $SO(4)=SU(2)_L\times SU(2)_R$, were it not
for the redefinition (\ref{zredef}), which breaks $SU(2)$ to its Cartan subalgebra, the action with $e^{+i\a}$ and $e^{-i\a}$ on the diagonal, 
which is identified with $U(1)_\pm$, since $x^+$ takes the role of the $\a$ parameter in (\ref{zredef}). 

In this case, we find for the lightcone momenta on the pp wave, in terms of the anomalous dimension 
$\Delta$ of the field theory and the charge $J_\psi$ associated with motion on $\psi$ in the gravitational background,
\begin{align}
 2p^- & = -p_+ = i\d_{x^+} = \sqrt{2} i \d_{\tilde{x}^+} = i \left(\d_t +  \frac{2}{3}  \d_\psi \right) = \Delta -  \frac{2}{3}  J_\psi\cr
 2p^+ & = -p_- = i\d_{x^-} = \frac{1}{\sqrt{2}} i \d_{\tilde{x}^-} = \frac{i}{R^2} \left(\d_t -  \frac{2}{3}  \d_\psi \right) = \frac{1}{R^2}\left(\Delta +  \frac{2}{3} 
  J_\psi\right).\label{mome}
\end{align}

In the field theory, we will identify $\frac{2}{3}J_\psi$ with $J$, the (large) charge of operators. In the picture where $\Delta [\phi^i]=1/2$
(as in the classical case), $J[\phi^i]=1/2$.

\subsubsection{Closed string quantization on the pp wave}

Using the definitions of \cite{Michelson:2002ps}, the Green-Schwarz action for the type IIA closed string on the pp-wave above is
found to be (like in the ABJM case) 
\be 
\begin{split}
 S = \frac{1}{4\pi \a'} & \int \dd t \int_0^{2\pi \a' p^+} \dd\s \left\{\sum_{i=1}^8 \left[ (\dot{X}^i)^2 - (X'^i)^2 \right] - \sum_{A=1}^4 (X^A)^2  \right.\\
 & \left. - \frac{1}{4}\sum_{B=5}^8 (X^B)^2 - 2i\bar{\Theta} \Gamma^- \left[ \d_\tau + \Gamma^{11}\d_\s + \frac{e^{-\phi_0}}{4\sqrt{2}}
 \left( \Gamma^1 \Gamma^{11} + 3 \Gamma^{234}\right)\right] \Theta \right\}
\end{split}
\ee
where we consider the identification $X^1=u$, $(y_1,y_2,y_3)\equiv X^A = (X^2, X^3,X^4)$ and finally $(w_i,\bar{w}_i)\equiv X^B 
= (X^5, X^6, X^7, X^8)$ (see \cite{Gaiotto:2008cg, Cardona:2014ora}). Therefore, the light-cone Hamiltonian for the closed string 
on the pp-wave is (we could rescale $x^+$ by $\mu$ as usual, and then $H$ would have a common $\mu$ factor to give it dimensions, 
but we keep it like this for ease of comparison with the field theory)
\be
H= \sum_{n=-\infty}^{\infty}  \left\{
 \sum_{A=1}^{4} N_n^{(A)}\sqrt{1+ \frac{n^2}{(\a' p^+)^2}} + \sum_{B=5}^{8} N_n^{(B)}\sqrt{\frac{1}{4} + \frac{n^2}{(\a' p^+)^2}}
\right\}\; .
\ee

If $n/(\a' p^+)\ll 1$, we find four modes, corresponding to $X^1=u,(X^2,X^3,X^4)=(y_1,y_2,y_3)$, with energies
\be
E^A\simeq 1+\frac{1}{2}\frac{n^2}{(\a'p^+)^2}\;,\label{ea}
\ee
and four modes, corresponding to $(X^5,X^6,X^7,X^8)$, with energies
\be
E^B\simeq \frac{1}{2}+\frac{n^2}{(\a'p^+)^2}.\label{eb}
\ee

\subsection{Useful pp-waves for (closed and) open strings on D-branes:\\ motion on $\sigma$}

In this subsection, we consider the Penrose limit  of the GJV solution near a null geodesic moving in the coordinate $\s$, 
with $\a=\pi/2$, $\l=\pi/4$, $\theta=\pi/2$ and $\r=0$ (also $\phi=0$, $\psi=0$), as we discussed for case 2 in the first subsection.
As we mentioned, this limit will be more useful for open strings on D-branes in the geometry than for closed strings. 
The D4-brane giant graviton wraps $\mathbb{CP}^2$, i.e. coordinates $(\lambda,\theta,\phi,\sigma)$, and moves in the $\psi$ 
direction, like the closed strings. The open strings must move in one of the isometry directions parallel to the D4-brane, so the 
$\sigma$ direction fits the bill. 
 
In order to boost on $\sigma$, we consider the transformation $\s \to\tilde{\s}= \frac{\sqrt{3}}{4} \s$ and the definitions
\be 
t = \frac{1}{\sqrt{2}} \left( x^+ + \frac{x^-}{R^2} \right), \quad \tilde{\s} =  \frac{1}{\sqrt{2}} \left( x^+ - \frac{x^-}{R^2} \right) \, , 
\ee
where, as before, $R=e^{\frac{\phi_0}{4}}L$. To find the required Penrose limit, we consider the expansion near the null geodesic above, namely 
\begin{align}
 \r & = \frac{\tilde{\r}}{\sqrt{2}R}\; , \quad \a = \frac{\pi}{2} + \frac{v}{\sqrt{3} R}\; , \quad \l = \frac{\pi}{4} + \frac{x}{\sqrt{6}R}\; ,\\ 
 \phi & = \frac{2}{\sqrt{3}} \frac{y}{R} \; , \quad \theta = \frac{\pi}{2} +\frac{2}{\sqrt{3}} \frac{z}{R} \; , \quad \psi =\frac{\sqrt{2}}{3} \frac{w}{R}\; .
\end{align}

Moreover, for a nontrivial Penrose limit in the direction $\s$, we make use of the fact that there is a freedom 
in the definition of $\omega$ in the metric, so that we consider the modification
\be 
\omega \to \omega' = \omega - \frac{1}{4}\dd \s\; ,
\ee
which obeys $\dd \omega = \dd \omega'$, as we explained when analyzing case 2 for the Penrose limit. As a result, we can then write $\eta = \dd \psi + \omega'$. 

Taking the limit $R\rightarrow \infty$, the metric becomes a pp-wave, 
\be
\label{sigma_pp}
\begin{split}
\dd s^2_{pp}  = & -4 \dd x^+ \dd x^- +  \dd \tilde{\rho}^2 + \tilde{\rho}^2 \dd \Omega_2^2  +   \dd v^2 + \dd w^2 +  \dd x^2 + \dd y^2 + \dd z^2 \\
& + 2\sqrt{2}\left(\frac{x}{2} \dd w - \frac{z}{\sqrt{3}} \dd y \right) \dd x^+   - \left( \frac{\tilde{\r}^2}{2} + \frac{2v^2}{3}+ \frac{x^2}{6}
\right) (\dd x^+)^2\;,
\end{split}
\ee
and the fields supporting the geometry may be expressed as follows: 
\be
\begin{split}
e^{\phi} &=  \sqrt{2} e^{\phi_0}, \quad B = - \frac{1}{\sqrt{3}} \dd v \wedge \dd w + \frac{1}{\sqrt{6}} v \dd x \wedge \dd x^+ - \frac{1}{\sqrt{6}} x \dd v \wedge \dd x^+\, ,  \\
\widetilde{F}_4 &= \frac{3}{2 e^{\phi_0}} \dd x^+ \wedge \tilde{\rho}^2 \, \dd \tilde{\rho} \wedge \textrm{vol}(S^2)  + \frac{1}{\sqrt{3} e^{\phi_0}} \dd x^+ \wedge \dd x \wedge \dd z \wedge \dd y.  
\end{split}
\ee
At this stage, we can check that the Einstein equation is satisfied. To aid the reader, we record that the only nonzero component of the Ricci tensor is $R_{++} = \frac{35}{12}$. Substituting this back into the Einstein equation (\ref{Einstein}), we find that it is satisfied.

In contrast to the solution presented in the last section, which preserved twenty-four supersymmetries, here it is easy to confirm that only sixteen supersymmetries are preserved. To see this, recall the dilatino variation of type IIA supergravity (see \cite{Itsios:2012dc} for notation)
\be
\delta \lambda = \frac{1}{2} \slashed{\partial} \phi \epsilon - \frac{1}{24} \slashed{H} \sigma^3 \epsilon + \frac{1}{8} e^{\phi} \left[ \frac{3}{2} \slashed{F}_2 ( i \sigma^2) + \frac{1}{24} \slashed{F}_4 \sigma^1 \right] \epsilon.  
\ee
Plugging in the solution, while ignoring the obvious projection condition that preserves sixteen supersymmetries, i. e. $\Gamma^+ \epsilon = 0$, we encounter the projection condition: 
\be
\frac{3 \sqrt{3}}{4} \Gamma^{v x \tilde{\rho} \vartheta \varphi} \epsilon + \frac{1}{2} \Gamma^{v z y} \epsilon = i \sigma^2 \epsilon. 
\ee
It is easy to convince oneself that this is not a good projection condition and does not permit any supernumeracy Killing spinors, namely those beyond the obvious sixteen.

To tidy up and bring the pp-wave solution to the standard Brinkmann form, we make the coordinate transformation
\be
x^-\rightarrow x^-+\frac{\sqrt{2}}{8} xw -\frac{\sqrt{2}}{4\sqrt{3}}zy\;,
\ee
in order to change the $g_{+ \mu}$, $\mu \neq +$, term in the metric,
\be
\begin{split}
\dd s^2_{pp} &  =  -4 \dd x^+ \dd x^- +  \dd \tilde{\rho}^2 + \tilde{\rho}^2 \dd \Omega_2^2  +   \dd v^2 + \dd w^2 +  \dd x^2 + \dd y^2 + \dd z^2 \\
& + 2\sqrt{2}\left(\frac{1}{4}(x \dd w-w\dd x) - \frac{1}{2\sqrt{3}} (z \dd y-y \dd z) \right) \dd x^+   - \left( \frac{\tilde{\r}^2}{2} + \frac{2v^2}{3}
+ \frac{x^2}{6}\right) (\dd x^+)^2.
\end{split}
\ee
Defining complex coordinates
\be 
z_1  = x + i w\;,\;\;\;
z_2  = z + i y\; ,
\ee
we obtain
\begin{align}
 \dd s^2_{pp}  & =  -4 \dd x^+ \dd x^- +  \dd \tilde{\rho}^2 + \tilde{\rho}^2 \dd \Omega_2^2  +   \dd v^2 + \sum_{i=1}^2 \dd z_i \dd \bar{z}_i 
 + \frac{\sqrt{2}i}{4}\left(z_1 \dd \bar{z}_1 - \bar{z}_1  \dd z_1\right)\dd x^+\nn\\
& -\frac{i}{\sqrt{6}}\left(z_2 \dd \bar{z}_2 - \bar{z}_2  \dd z_2\right)\dd x^+   - \left( \frac{\tilde{\r}^2}{2} + \frac{2v^2}{3}
+ \frac{\mathrm{Re}(z_1)^2}{6}\right)(\dd x^+)^2.\label{pp01}
\end{align}
To bring the metric to Brinkmann form (\ref{Brinkmann}), we consider the coordinate transformations, 
\be
\label{replacement}
z_1= e^{-\frac{i\sqrt{2}x^+}{4}}w_1\;,\;\;\;
z_2=e^{\frac{i x^+}{\sqrt{6}}}w_2\;,
\ee
after which the metric takes the form we want: 
\be
\begin{split}
\label{pp-metric}
\dd s^2_{pp}  = & -4 \dd x^+ \dd x^- +  \sum_{i=1}^3 \dd x_i^2  + \sum_{k=4}^8 \dd y_k^2 \cr
&   - \left( \frac{1}{2}\sum_{i=1}^3 x_i^2 + \frac{2 y_4^2}{3}+\frac{y_5^2+y_6^2}{8}+\frac{y_7^2+y_8^2}{6}\right) (\dd x^+)^2  \\
&-\frac{1}{6}\left[y_5  \cos \left( \frac{\sqrt{2}x^+}{4} \right)+y_6 \sin \left( \frac{\sqrt{2} x^+}{4} \right) \right]^2(\dd x^+)^2\; .
\end{split}
\ee
where we have introduced $x_i, i = 1, 2, 3$ to parametrize the $\tilde{\rho}$ directions and have employed the redefinitions: $v=y_4$, Re$(w_1)= y_5$, Im$(w_1)= y_6$, Re$(w_2)= y_7$, Im$(w_2)=y_8$. Rewriting the remainder of the solution in terms of Brinkmann coordinates, one finds: 
\bea
B &=& -\sqrt{\frac{2}{3}}\left[y_5 \cos \left( \frac{\sqrt{2} x^+}{4} \right)+y_6 \sin \left( \frac{\sqrt{2} x^+}{4} \right) \right] \dd y_4\wedge \dd x^+ \;, \label{b-field} \cr
\widetilde{F}_4 &=& \frac{3 e^{-\phi_0}}{2 } \dd x^+ \wedge \dd x_1 \wedge \dd x_2 \wedge \dd x_3\cr
&&+ \frac{e^{-\phi_0}}{\sqrt{3}} \dd x^+ \wedge \left[   \cos \left(\frac{\sqrt{2}x^+}{4} \right) \dd y_5 + \sin \left( \frac{\sqrt{2} x^+}{4} \right)  \dd y_6 \right] \wedge \dd y_7 \wedge \dd y_8  \; .
\eea
where we have employed gauge symmetry to bring $B$ to the above form. 

Given the fact that the energy is  $E=i\d_t$, which in the dual field theory corresponds to the conformal dimension $\Delta$, 
and that the angular momentum for motion in $\sigma$ is $J_\s=-i\d_\s$, the lightcone momenta on the pp-wave are written in terms of 
$\Delta$ and $J_\sigma$ as
\bse
\begin{align}
2p^- & = -p_+=i \d_+ = \frac{1}{\sqrt{2}}\left(\Delta - \frac{4}{\sqrt{3}} J_\s \right)\\ 
2p^+ & = -p_-=i \d_- = \frac{1}{\sqrt{2}R^2}\left(\Delta + \frac{4}{\sqrt{3}} J_\s \right)\; .
\end{align}
\ese

\subsection{Useful pp waves for (closed and) open strings on D-branes:\\ motion in $\phi$}

For open strings on D4-brane giant gravitons we have an alternative: take the Penrose limit on null geodesics moving in the isometry direction 
$\phi$ instead of $\sigma$, around $\theta=\pi/2, 
\lambda=\pi/2, \a=\pi/2, \rho=0$, i.e. case 3 in our general analysis.

We expand these coordinates as follows: 
\be
\rho = \frac{\tilde{\rho}}{\sqrt{2} R}, \quad \alpha = \frac{\pi}{2} + \frac{v}{\sqrt{3} R}, \quad \lambda = \frac{\pi}{2} 
+\frac{x}{\sqrt{6} R}, \quad  \theta = \frac{\pi}{2} + \sqrt{\frac{2}{3}}\frac{ z}{ R}. 
\ee
In addition, we also redefine 
\bea
 \sigma &=& 2 \tilde{\sigma},  \quad  \psi = \frac{\sqrt{2}}{3 R} \tilde{\psi} - \tilde{\sigma}, \quad \phi = \frac{2}{\sqrt{3}} \tilde{\phi}, \nn \\
t &=& \frac{1}{2} \left( x^+ + \frac{x^-}{R^2}  \right), \quad \tilde{\phi} = \frac{1}{2} \left( x^+ - \frac{x^-}{R^2}  \right), 
\eea
so that the metric in the pp-wave limit ($R  = e^{\frac{\phi_0}{4}} L \rightarrow \infty$) becomes, 
\bea
 \dd s^2 &=& -2 \dd x^+ \dd x^-  - z \dd \tilde{\psi} \dd x^+ + \dd \tilde{\rho}^2 + \tilde{\rho}^2 \dd s^2(S^2) 
 + \dd v^2 + \dd x^2 + x^2 \dd \tilde{\sigma}^2 + \dd z^2 + \dd \tilde{\psi}^2 \nn \\
&-&  \left[ \frac{1}{3} v^2 + \frac{1}{4} \tilde{\rho}^2 + \frac{1}{12} x^2 + \frac{1}{12} z^2 \right] (\dd x^+)^2 . 
\eea
We also perform the same limiting procedure on the rest of the solution:
\bea
e^{\Phi} &=& \sqrt{2} e^{\phi_0}, \nn \\
\widetilde{F}_4 &=& \frac{3}{2 \sqrt{2} e^{\phi_0}} \dd x^+ \wedge \tilde{\rho}^2 \dd \tilde{\rho} \wedge \textrm{vol}(S^2) 
+  \frac{1}{\sqrt{6} e^{\phi_0}}  \dd x^+ \wedge x \dd x \wedge \dd z \wedge \dd \tilde{\sigma}, \nn \\
H_3 &=& \frac{1}{\sqrt{3}} \dd x^+ \wedge \dd z \wedge \dd v. 
\eea
To make sure that everything is correct, one can quickly check that the Einstein equation (\ref{Einstein}) is satisfied. 
Note that $R_{++} = \frac{35}{24}$ is the only nonzero component of the Ricci tensor. 

Having established that we have taken the pp-wave limit correctly, one can bring it to the Brinkmann form by redefining \footnote{Broken
 down in terms of steps this transformation involves a shift $x^{-} \rightarrow x^{-} - \frac{1}{4} z \tilde{\psi}$ and the redefinition 
 $Z = e^{\frac{i}{4} x^{+}} W$, where $Z = z + i \tilde{\psi}$ and $W = w_1 + i w_2$.}: 
\bea
x^{-} &\rightarrow& x^{-} - \frac{1}{8} \sin \frac{x^+}{2} (w_1^2 -w_2^2)- \frac{1}{4} \cos \frac{x^+}{2} w_1 w_2 , \nn \\
z &\rightarrow& \cos \frac{x^+}{4} w_1 - \sin \frac{x^+}{4} w_2, \nn \\
\tilde{\psi} &\rightarrow& \sin \frac{x^+}{4} w_1 + \cos \frac{x^+}{4} w_2. 
\eea
The end result may be expressed as:
\bea
 \dd s^2 &=& -2 \dd x^+ \dd x^-  + \dd \rho^2 + \rho^2 \dd s^2(S^2) + \dd v^2 + \dd x^2 + x^2 \dd \sigma^2 + \dd w_1^2 + \dd w_2^2 \nn \\
&-&  \left[ \frac{1}{3} v^2 + \frac{1}{4} \rho^2 + \frac{1}{12} x^2 + \frac{1}{12} \left(  \cos \frac{x^+}{4} w_1 
- \sin \frac{x^{+}}{4} w_2\right)^2 + \frac{1}{16} (w_1^2 + w_2^2) \right] (\dd x^+)^2, \nn \\
e^{\phi} &=& \sqrt{2} e^{\phi_0}, \nn \\
\widetilde{F}_4 &=& \frac{3}{2 \sqrt{2} e^{\phi_0}} \dd x^+ \wedge \rho^2 \dd \rho \wedge \textrm{vol}(S^2)
 +  \frac{1}{\sqrt{6} e^{\phi_0}}  \dd x^+ \wedge x \dd x \wedge \dd z \wedge \dd \sigma, \nn \\
H_3 &=& \frac{1}{\sqrt{3}} \dd x^+ \wedge \left( \cos \frac{x^+}{4} \dd w_1 - \sin \frac{x^+}{4} \dd w_2 \right) \wedge \dd v\;,
\eea
where we have dropped tildes. 

It is easy to see from supersymmetry analysis, the details of which we omit, that only sixteen supersymmetries are preserved. 

\subsection{Useful pp waves for (closed and) open strings on D-branes:\\ motion in $\phi+\sigma$}

Yet another alternative for open strings on D4-brane giant gravitons is to take the Penrose limit on some 
combination of $\sigma$ and $\phi$. We can choose it for simplicity to 
be $\sigma+\phi$, as in case 4 of our general analysis. 

Then we must expand around the null geodesic with $\lambda=\pi/4$, $\a=\pi/2$, 
$\theta=0$, $\rho=0$, and we can choose $\psi=0$. Thus we expand the variables as
\bea
\theta=\frac{ 2 z}{\sqrt{3} R} && \lambda=\frac{\pi}{4}+\frac{x}{ \sqrt{6} R}, \;\;\;\; \a=\frac{\pi}{2}+\frac{v}{\sqrt{3} R}\cr
\psi=\frac{\sqrt{2} w}{3 R} && \rho=\frac{\tilde \rho}{\sqrt{2}R}.
\eea
In addition, we redefine 
\bea
\sigma'= 2 \sqrt{\frac{2}{3}}  \tilde \sigma, \quad 
\phi'=\sqrt{2}\tilde \phi, \quad t = \frac{1}{\sqrt{2}} \left( x^+ + \frac{x^-}{R^2} \right), \quad \tilde{\sigma} =  \frac{1}{\sqrt{2}} \left( x^+ - \frac{x^-}{R^2} \right), 
\eea
to obtain the pp-wave metric, 
\bea
\dd s^2&=&-4\dd x^+\dd x^-+\dd \tilde \rho^2+\tilde \rho^2 \dd \Omega_2^2+\dd v^2+\dd w^2+\dd x^2+\dd z^2+z^2\dd \tilde \phi^2\cr
&&-(\dd x^+)^2\left[ \frac{\tilde \rho^2}{2}+\frac{2v^2}{3} +\frac{x^2}{6} \right]+ \dd x^+ \left[- \sqrt{\frac{2}{3}} z^2 \, \dd \tilde \phi+ \sqrt{2}  x \, \dd w\right].
\eea
Taking the same limit on the rest of the solution, one finds: 
\bea
e^{\phi} &=& \sqrt{2} e^{\phi_0}, \quad B = \sqrt{\frac{2}{3}} v \dd x \wedge \dd x^+, \nn \\
\tilde{F}_4 &=& \frac{3}{2 e^{\phi_0}} \dd x^+ \wedge \tilde{\rho}^2 \dd \tilde{\rho} \wedge \textrm{vol}(S^2) 
 - \frac{z}{\sqrt{3} e^{\phi_0}} \dd x^+ \wedge \dd x \wedge \dd z \wedge \dd \tilde{\phi}, 
\eea
where we have used gauge symmetry to drop a total derivative from the $B$-field. Using $R_{++}  = \frac{35}{12}$, the only 
nonzero component of the Ricci tensor, it is an easy exercise to convince oneself that the Einstein equation is satisfied. In fact, 
it is worth noting that the solution is modulo coordinate redefinitions and irrelevant signs \footnote{Note, the metric is essentially 
quadratic in coordinates, whereas the fluxes are linear, so via reflection, we can change the sign of the fluxes.}, the same as 
the pp-wave limit for motion on the $\sigma$ direction (\ref{sigma_pp}). To bring it to the same form, one can introduce cartesian 
coordinates $ z_1 = z \sin \tilde{\phi}, \quad z_2 = z \cos \tilde{\phi}$, in addition to shifting $x^- \rightarrow x^- + 1/(2 \sqrt{6}) z_1 z_2$. 
As a result, the rewriting of the pp-wave in terms of Brinkmann coordinates reduces to the previous analysis and we omit further details.

\subsection{Open string quantization on the pp-waves}\label{openquant}

 In this subsection we consider open string quantization making use of the pp-wave limits identified in the previous subsections. We will 
 encounter difficulties in solving the entire system, but we will find a common sector that can be solved for all the pp-waves we 
 have identified. As a result, we focus our attention on the pp-wave obtained for motion in the $\sigma$ direction.  

The bosonic sector of type IIA string theory, in the presence of a $B$-field, is described by the Polyakov action 
\be 
S=-\frac{1}{4\pi\a'} \int \dd^2 \s \left( \eta^{ab}G_{\m\n}\d_a X^\m \d_b X^\n + \epsilon^{ab} B_{\m\n} \d_a X^\m \d_b X^\n \right)\; ,
\ee 
where we have used the conformal gauge $\sqrt{-h}h^{ab}=\eta^{ab}=\textrm{diag}(-1,1)$ and $\epsilon^{01}=1$. 
To avoid confusion, we write the worldsheet coordinates as $(\s^0, \s^1)$. 

For the pp-wave corresponding to motion on $\sigma$, inserting the pp-wave solution (\ref{pp-metric}) 
and (\ref{b-field}), and making a rescaling $x^+\to \m x^+$ and $x^-\to \m^{-1} x^-$, along with the identifications
$X_i = x_i, i = 1, 2, 3$, $Y_{i} = y_{i}, i = 4, 5, 6, 7, 8$, the string action on the pp-wave becomes
(in the light-cone gauge $ \m \, x^+=\sigma^0$)
\be 
\begin{split}
S_{pp} & =-\frac{1}{4\pi \a'}\int \dd \s^0 \int_0^{\pi\a'p^+} \dd\s^1 \left[
\eta^{ab}(\d_a X_i \d_b X_i + \d_a Y_k \d_b Y_k) + 
\m^2\left(\frac{X_i^2}{2} + \frac{2 Y_4^2}{3} \right. \right. \\
& \left. \left. + \frac{Y_5^2+Y_6^2}{8}+ \frac{Y_7^2 + Y_8^2}{6} +\frac{1}{6}\left[Y_5 \frac{\sqrt{2}\sigma^0}{4}
+Y_6 \sin \frac{\sqrt{2} \sigma^0}{4}\right]^2 \right)\right.\cr
&\left.+ 2\sqrt{\frac{2}{3}}\m \left[Y_5 \frac{\sqrt{2} \sigma^0}{4}+Y_6 \sin  \frac{\sqrt{2} \sigma^0}{4}\right] \d_1 Y_4 \right] \;.
\end{split}\label{stringpp}
\ee
The equations of motion read
\bse
\be
\begin{split}
 \d^2 X_i & -\frac{\m^2}{2} X_i = 0\ \ ,\ \  \d^2 Y_4  - \frac{2\m^2}{3} Y_4 + \sqrt{\frac{2}{3}}\m
 \left[(\d_1Y_5) \cos\frac{\sqrt{2}\sigma^0}{4}+(\d_1Y_6) \sin \frac{\sqrt{2} \sigma^0}{4}\right]  = 0\ \ ,\cr 
 \d^2 Y_5 & - \frac{\m^2}{8}Y_5 -\frac{\mu^2}{6}\cos\frac{\sqrt{2}\sigma^0}{4}
 \left[Y_5 \cos\frac{\sqrt{2}\sigma^0}{4}+Y_6 \sin \frac{\sqrt{2} \sigma^0}{4}\right]
 - \sqrt{\frac{2}{3}}\m \cos \frac{\sqrt{2} \sigma^0}{4}  \d_1 Y_4= 0\ ,\\
 \d^2 Y_6 & - \frac{\m^2}{8} Y_6 -\frac{\mu^2}{6}\sin\frac{\sqrt{2}\sigma^0}{4}
 \left[Y_5 \cos\frac{\sqrt{2}\sigma^0}{4}+Y_6 \sin \frac{\sqrt{2} \sigma^0}{4}\right]
 - \sqrt{\frac{2}{3}}\m \sin \frac{\sqrt{2} \sigma^0}{4}  \d_1 Y_4 = 0\ \ , \cr
   \d^2 Y_7 & - \frac{\m^2}{6}Y_7 = 0\ \ , \ \
 \d^2 Y_8 - \frac{\m^2}{6}Y_8 = 0\; ,
\end{split}
\ee
and the general boundary conditions are
\be
\begin{split}
& \d_1 X_i \delta X_i = 0 \;  (i=1,2,3)\; ,\\
& \d_1 Y_4 \delta Y_4 +\sqrt{\frac{2}{3}}  \left[Y_5 \cos\frac{\sqrt{2}\sigma^0}{4}+Y_6 \sin \frac{\sqrt{2} \sigma^0}{4}\right]  \delta Y_4 = 0 \;, \\ 
 & \d_1 Y_I \delta Y_I = 0\; , (I=5,6,7,8)\ \; .
\end{split}
\ee
\ese

We are interested in open strings attached to a D$4$-brane wrapping the $\mathbb{R}_t\times \mathbb{CP}^2$ space spanned by the coordinates 
$(t,\lambda,\theta,\phi,\sigma)$,  which become $(x^\pm, x, y,z)$ in the pp-wave limit, or after the redefinitions, 
$(x^\pm, y_7, y_8)$ and $\left[y_5 \cos\frac{\sqrt{2}\sigma^0}{4}+y_6 \sin \frac{\sqrt{2} \sigma^0}{4}\right]$ 
\cite{Berenstein:2002zw, Dabholkar:2002zc}. Therefore, we impose Neumann boundary conditions 
along these directions and Dirichlet boundary conditions in the remaining directions,
\be 
\begin{split}
\d_1 x^\pm & = \d_1 \left[Y_5 \cos\frac{\sqrt{2}\sigma^0}{4}+Y_6 \sin \frac{\sqrt{2} \sigma^0}{4}\right]  = \d_1 Y_7 = \d_1 Y_8 = 0\\ 
\delta X_i & = \delta Y_4 =  \delta\left[-Y_5 \sin\frac{\sqrt{2}\sigma^0}{4}+Y_6 \cos \frac{\sqrt{2} \sigma^0}{4}\right] = 0\; .
\end{split}
\ee
Note that consistency necessarily implies that $x^\pm$ satisfy Neumann boundary conditions. 

The coupled system of $Y_4,Y_5,Y_6$ is difficult to solve, but $X_i, Y_7,Y_8$ are simple. For them, we find 
\bse
\begin{align}
 X_i & = -\sqrt{2\a'} \sum_{n\neq 0 } \frac{\a^{(i)}_n}{\omega_n^{(i)}} \sin\left(\frac{n \s}{\a'p^+}\right) e^{-i\omega_n^{(i)} \tau}\; , \\
 Y_{I'} & = y_0^{(I')}\cos\left(\frac{\m \tau}{\sqrt{6}}\right) + \sqrt{6}\a' p_0^{(I')} \sin \left(\frac{\m\tau}{\sqrt{6}}\right) 
 + i \sqrt{2\a'} \sum_{n \neq 0}\frac{\a_n^{(I')}}{\omega_n^{(I')}} \cos \left(\frac{n \s}{\a'p^+}\right)  e^{-i \omega_n^{(I)} \tau}\; ,
\end{align}\label{xiyi}
\ese
for $I'=7,8$. The eigenenergies for these modes are
\be
\omega_n^{(i)}  = \sqrt{\frac{\m^2}{2} + \frac{n^2}{(\a'p^+)^2}}\ , \quad  \omega_n^{(I')} = \sqrt{\frac{\m^2}{6} + \frac{n^2}{(\a'p^+)^2}}, \;\; 
I'=7,8. \label{omegaopen}
\ee
In principle, the remaining equations can be solved numerically, but we postpone this to future work.

In fact, we can check that the part of the string action (\ref{stringpp}) involving $X_i, Y_{I'}$ is the same for all three pp-wave metrics, namely for motion on $\phi$, on 
$\sigma$, or on $\phi+\sigma$, and therefore the solutions (\ref{xiyi}) and the eigenenergies (\ref{omegaopen}) are the same for all. 
We will see that a puzzle arises when we try to construct an open string spin chain in the field theory: it has to correspond to 
one of the pp-waves, and all of them have the same energies (\ref{omegaopen}) for the modes (\ref{xiyi}). In spite of this, we cannot get the zeroth order 
spin chain energies to agree with (\ref{omegaopen}).

\section{Spin chains in the field theory IR limit}

As we have seen in section 2, in the IR limit of the field theory, if we choose the picture where we use the classical description of the theory,
we have states satisfying $[\phi_i,\phi_j] = 0$, $\forall ~i,j=1,2,3$, and for these states
the bosonic interaction Hamiltonian is given by (\ref{intpot}), which we repeat here for convenience
\be
H_{\rm int, 1}=\frac{4\pi^2}{k^2}\Tr\left([[\phi^{i\dagger},\phi^i],\phi^{k\dagger}][[\phi^{j\dagger},\phi^j],\phi^k]\right)\; .
\ee

\subsection{Closed string spin chain}

The symmetry of the theory is $SO(2,3)_{\rm conf.}\times SU(3)\times U(1)$, and in the field theory Penrose limit, which should be a large R-charge limit, we 
expect, based on the gravitational Penrose limit on the $\psi$ direction, 
the $U(1)$ to combine with the $SO(2)$ (scaling) part of the conformal $SO(2,3)$ to give 
$U(1)_\pm$, and the $SU(3)$ will break to $SU(2)_L$. We expect also an extra $U(1)_u$ symmetry. 

This breaking
coincides with the idea of picking out, as usual, a special complex field charged under the $U(1)$, let us call it $Z$, 
among the $\phi_i$, $i=1,2,3$. We call the remaining complex scalars  $\phi_m$, $m=1,2$. Then the bosonic interaction Hamiltonian
in the IR is
\bea
H_{\rm int}&=&\frac{4\pi^2}{k^2}\Tr\left([[\bar Z,Z],\bar Z][[\bar Z,Z],Z]+[[\bar\phi^m,\phi^m],\bar Z][[\bar\phi^n,\phi^n],Z]\right.\cr
&&\left.+[[\bar Z,Z],\bar\phi^m][[\bar Z,Z],\phi^m]+[[\bar\phi^m,\phi^m],\phi^p][[\bar\phi^n,\phi^n],\phi^p]\right.\cr
&&\left.+[[\bar Z,Z],\bar\phi^n][[\bar \phi^m,\phi^m],\phi^n]+[[\bar\phi^m,\phi^n],\bar\phi^n][[\bar Z,Z],\phi^n]\right.\cr
&&\left.+[[\bar Z,Z],\bar Z][[\bar\phi^m,\phi^m],Z]+[[\bar Z,Z],Z][[\bar\phi^m,\phi^m],\bar Z]\right).
\eea

We want to construct the closed string spin chain, whose vacuum must have charge $J$, corresponding to momentum $p^+$ on the pp-wave. 
We have to define the charge of $Z$ (the unit of charge), and we define it to be such that (in our picture using classical dimensions)
$\Delta-J=0$ for $Z$, i.e. $J=1/2$. 
In (\ref{mome}), we saw that $2p^-=\Delta-(2/3)J_\psi$, and we said that we identify $(2/3)J_\psi$ with $J$. Moreover, we want to have 
zero energy for states made up of only $J$'s (vacuum states), so $\Delta[Z]=J[Z]=1/2$. 

This then implies also $J[\phi^m]=J[\bar\phi^m,]=0$, and $J[\bar Z]=-1/2$, so all in all, we have the table 
\begin{center}
	\begin{tabular}{c|cccccc}
	                 & $Z$ & $\bar Z$ & $\phi^m$ & $\bar{\phi}^m$  & $A_\m$ & $D_\m$\\ \hline\hline
	$\Delta$    & $1/2$ & $1/2$ & $1/2$ &$1/2$  & $1$ & $1$\\
	$J$            & $1/2$ & $-1/2$ & $0$ & $0$ & $0$ & $0$\\ \hline
	$\Delta-J$ & $0$ & $1$ & $1/2$ & $1/2$ & $1$	& $1$
	\end{tabular}
\end{center}
Then the unique object of $\Delta - J=0$ (corresponding to $E=p^-=0$) and $J=1/2$ (corresponding to $p^+=1/2$) is $Z$, so the 
vacuum has to be made from it only. At $\Delta-J=1/2$ we have 
both $\phi^m$ and $\bar \phi^m$, $m=1,2$, which have $J=0$. At $\Delta-J=1$ we have $\bar Z$ and $D_a$, $a=1,2,3 $ (covariant derivatives 
in the field theory direction). 

We can use the same vacuum as in the BMN\cite{Berenstein:2002jq}  
case, since we want a vacuum with $J$ units of $p^+$, i.e.
\be
|0,p^+\rangle\sim\Tr[Z^J]\;,
\ee
or more precisely
\be 
|0, p^+\rangle \leftrightarrow \frac{1}{\sqrt{J} N^{J/2}}\Tr[Z^J].
\ee

Introducing oscillators on top of them is trickier, since now (in the ``classical" picture) $[\phi_i,\phi_j]=0$ (but $[\phi^i,\bar \phi^j]\neq 0$). 
We can imagine introducing $\bar \phi^m$ as oscillators, since they do not commute with $Z$, though it is less clear what to do about $\phi^m$, since they do.

Another possibility, since now also $[\phi^m,\bar\phi^m]$ has nonzero commutator with $Z$, we can put insertions of $[\phi^m,\bar\phi^m]$ instead. 
In that case, for instance a nontrivial insertion would be
\be
a_n^{\dagger m}|0,p^+\rangle\sim \sum_{l=0}^{J-1}e^{\frac{2\pi i n l}{J}}\Tr[Z^l [\bar\phi^m,\phi^m] Z^{J-l}].
\ee
In total, we could insert the 3+4+1=8 oscillators of $\Delta-J=1$, 
\be
[Z,\bar Z]; \;\;\; [\phi^m,\bar \phi^n];\;\;\; D_a.
\ee
But this has the disadvantage that the classical dimension (at $\lambda=g^2_{YM}N=0$) would be $\Delta-J=1$ for all the oscillators, yet we have 
seen that half the closed string oscillators in the pp-wave in $\psi$ have energy 1 at $n=0$ and half have energy 1/2.  

We can easily see that the gravity dual picture is instead matched by the simplest possibility, namely the same one as in the 3+1 dimensional ${\cal N}=4$
SYM case. Namely, the oscillators are $\phi^m, \bar \phi^m $, $m=1,2$, of $\Delta-J=1/2$, corresponding to the 4 modes of 
energy 1/2 at $\lambda=0$, $n=0$ on the pp-wave (modes $B$),  and $D_a$, $a=1,2,3$ and $\bar Z$, both of $\Delta-J=1$, corresponding to 
the 4 modes (modes $A$) of energy 1 at $\lambda=0$, $n=0$ on the pp-wave. Defining then a generic insertion as
\be
\Phi_M=\{\phi^m,\bar\phi^m,\bar Z,D_a\}\;,
\ee
their insertion inside the trace corresponds to a string oscillator as usual, 
\be
a_n^{\dagger M}|0,p^+\rangle\sim \sum_{l=0}^{J-1}e^{\frac{2\pi i n l}{J}}\Tr[Z^l \Phi^M Z^{J-l}].
\ee

Note then that if $\Phi$ ``hops" to the right when acting on the state with the Hamiltonian, it acquires a $e^{ip}$ factor ($p=2\pi n/J$), and if it ``hops"
to the left, it acquires a $e^{-ip}$ factor, etc.

If we would consider commutator insertions, 
then among the terms in $H_{\rm int}$, the first and the second give the interaction of $[Z,\bar Z]$ and $[\phi^m,\bar\phi^m]$, and the third, fourth, fifth and sixth  
will not contribute {\em in the dilute gas approximation}, since at the planar level they would need to have two oscillators next to each other (at the 
same site) to contribute. As a result, the terms that contribute are 
\be
H_{\rm int, dilute}=\frac{4\pi^2}{k^2}\left\{\Tr\left([[\bar Z,Z],\bar Z],[[\bar Z,Z],Z]\right)+\Tr\left([[\bar\phi^m,\phi^m],\bar Z][[\bar\phi^m,\phi^m],Z]\right)\right\}.
\ee
However, there are also the mixing terms on the last line (seventh and eight).

This in effect is the same kind of interaction as in ${\cal N}=4$ SYM, once we replace the oscillators $\phi^m$ by $[\bar Z,Z]$
or $[\bar \phi^m,\phi^m]$, so the result would be the same. However, as we said above, already at tree level we get a different 
result from the gravity dual, so we need to consider instead just the usual insertions  of 3+1 dimensional ${\cal N}=4$ SYM.

\subsection{Closed string eigenenergies and Hamiltonians}

Consider then bosonic $\bar\phi^m,\phi^m$, $D_a$ and $\bar Z$ insertions. 
For $\bar Z$ insertion,  the relevant part of the interaction Hamiltonian is 
\be
\Tr([[\bar Z,Z],\bar Z][[\bar Z,Z],Z])=\Tr[3\bar Z Z \bar Z^2 Z^2+3Z\bar Z Z^2\bar Z^2-4\bar Z Z \bar Z Z \bar Z Z -2 Z^3 \bar Z^3].
\ee

\subsubsection{Commutator $[\bar Z,Z]$ insertion}

To set up the procedure, let us start with the simplest case (though we showed it doesn't match the gravity dual) of commutator 
insertion. Then the above term in the interaction Hamiltonian is considered
as the interaction for the commutator, and we rewrite it as 
\be 
\begin{split}
\Tr \left( [[\bar Z, Z],\bar Z][[\bar Z, Z], Z] \right) =\Tr & ( [\bar Z, Z] \bar Z [\bar Z, Z] Z -  [\bar Z, Z] \bar Z Z [\bar Z, Z] - \\
& \bar Z[\bar Z, Z]  [\bar Z, Z] Z + \bar Z  [\bar Z, Z]  Z [\bar Z, Z] )\, .
\end{split}
\ee

We write the planar diagrams implied by this interaction {\em in the dilute gas approximation} (dual to the pp-wave limit)
\begin{itemize}
		\item The interaction $\Tr( [\bar Z, Z] \bar Z [\bar Z, Z] Z)$ gives the diagram in Fig.\ref{comm}.
		
\begin{figure}[h]
	\begin{center}
  \includegraphics[width=5.0cm]{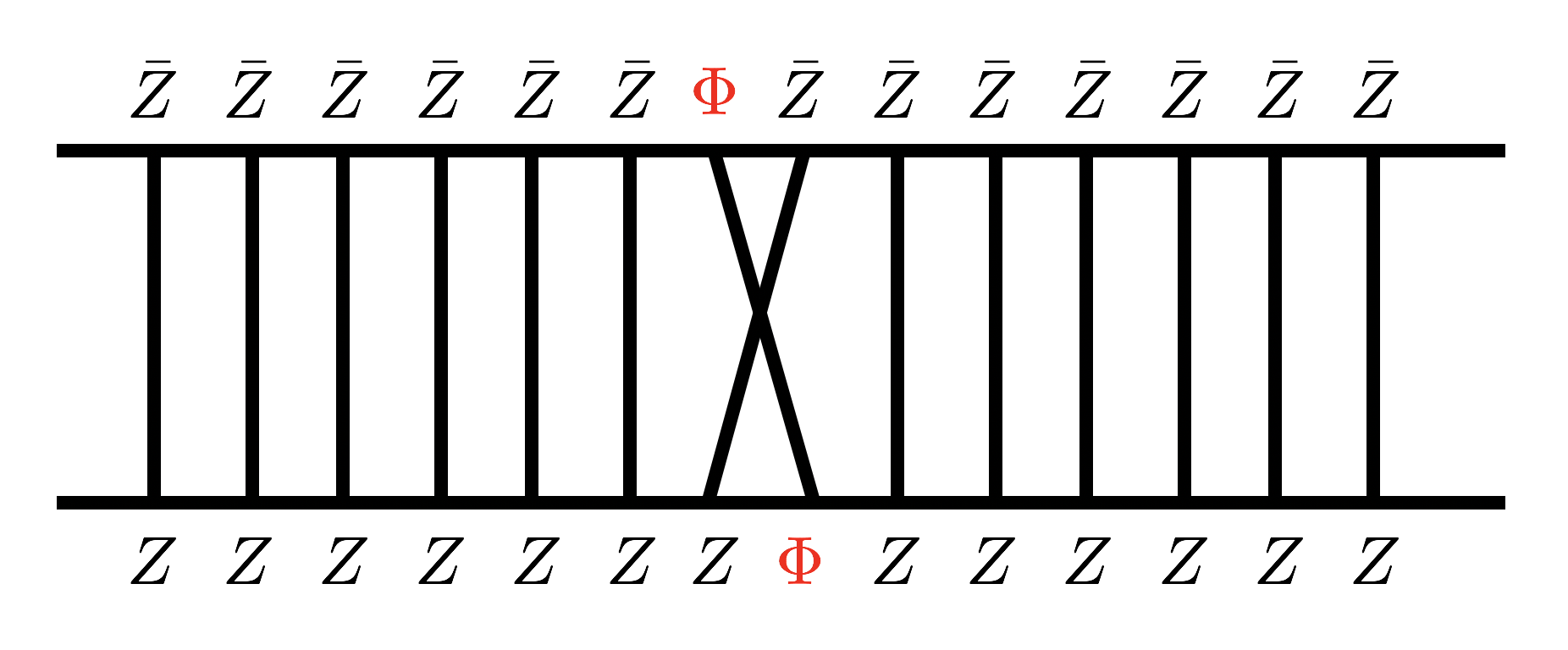}
	\end{center}\caption{Commutator interaction.}\label{comm}
\end{figure}

that we write as  $\left( \begin{smallmatrix} {\color{red} \Phi}&\bar Z\\ Z& {\color{red} \Phi} \end{smallmatrix} \right)$, 
where $\Phi=[\bar Z, Z]$. The coefficient for this diagram is $1$ and phase $e^{ip}$. Using the cyclicity of the trace, this interaction 
is equivalent to $\Tr(\bar Z [\bar Z, Z] Z [\bar Z, Z])$, which gives the diagram 
$\left( \begin{smallmatrix}\bar Z &  {\color{red} \Phi}\\ {\color{red} \Phi} & Z \end{smallmatrix} \right)$. 
The coefficient for this diagram is $1$ and phase the phase associated with it is $e^{-ip}$.

Note that there are vertices with two insertions, e.g. $\left( \begin{smallmatrix}{ \color{red} Z} &  {\color{red} \Phi}\\ {\color{red} \Phi} & 
{ \color{red} \bar Z}  \end{smallmatrix} \right)$ and $\left( \begin{smallmatrix}{ \color{red} \Phi} &  {\color{red} Z}\\ {\color{red} \bar Z} & 
{ \color{red} \Phi}  \end{smallmatrix} \right)$,   but in the dilute gas approximation we neglect these, since they are subleading.

\item The interaction $\Tr( [\bar Z, Z] \bar Z Z [\bar Z, Z])$ gives the diagram $\left( \begin{smallmatrix}  {\color{red} \Phi} & 
\bar Z\\ {\color{red} \Phi} & Z \end{smallmatrix} \right)$, with coefficient $-1$.

\item The interaction $\Tr(\bar Z [\bar Z, Z] [\bar Z, Z] Z)$ gives the diagram $\left( \begin{smallmatrix}  \bar Z & {\color{red} \Phi}\\  
Z & {\color{red} \Phi} \end{smallmatrix} \right)$, with coefficient $-1$.
\end{itemize}

Then the total factor associated with these planar diagrams is
\be 
f_{[\bar Z,Z]}=-2+(e^{ip}+e^{-ip})=-4\sin^2\frac{p}{2}.\label{fbarzz}
\ee
This would be the factor arising from the expansion in $\lambda$ of the BMN eigenenergies. 

It is associated with a term in the effective 
discretized string Hamiltonian obtained from rewriting the diagrams above in terms of ``fields at site $j$" $\phi_j$, as 
\be
-\sum_j(2\phi_j^2-2\phi_j \phi_{j+1})=-\sum_j(\phi_j-\phi_{j+1})^2\sim -\phi'^2.
\ee
This is just (minus) the ${\cal N}=4$ SYM result. 

\subsubsection{Insertion of $\bar Z$}

Now we move to the physical case, of insertion of fundamental fields, and we start with a $\bar Z$ insertion. 
In this case, we need to remember that the space of physical states is defined by $[\phi_i,\phi_j]=0$, so we can 
freely commute $\phi_m$ past $Z$'s, and $\bar \phi_m$ past $\bar Z$. We then have the following 
diagrams:

\begin{itemize}
\item The interaction term $\Tr( \bar Z  Z \bar Z Z \bar Z Z)$ gives two diagrams $\left( \begin{smallmatrix} \bar Z & {\color{red}Z} 
& \bar Z\\ Z&{\color{red} \bar Z} & Z \end{smallmatrix} \right)$ and $\left( \begin{smallmatrix} {\color{red} Z} & {\color{red}\bar Z} 
& Z\\ {\color{red}\bar Z} & Z & {\color{red} \bar Z} \end{smallmatrix} \right)$. 
		
Both diagrams are planar, since the vertex is sextic, but again we neglect the second diagram since we are in the dilute gas approximation. 
The coefficient  of the first diagram is $-4$ and its phase is trivial (=1).

\item The interaction term $\Tr( \bar Z Z \bar Z^2 Z^2)$ gives three relevant diagrams $\left( \begin{smallmatrix} \bar Z & {\color{red}Z} 
& \bar Z\\ Z& Z & {\color{red} \bar Z} \end{smallmatrix} \right)$, 
$\left( \begin{smallmatrix} {\color{red} Z} & \bar Z & \bar Z\\{\color{red} \bar Z} & Z &  Z \end{smallmatrix} \right)$ and 
$\left( \begin{smallmatrix} \bar Z & \bar Z &  {\color{red} Z} \\ Z &  {\color{red} \bar Z} & Z \end{smallmatrix} \right)$. 
The coefficient for each diagram is $1$, since the total coefficient is $3$, and their phases are, respectively $e^{ip}$, $1$ and  $e^{-ip}$.

\item The interaction term $\Tr( Z \bar Z Z^2 \bar Z^2)$ gives three relevant diagrams $\left( \begin{smallmatrix} {\color{red} Z} & \bar Z 
& \bar Z\\ Z& {\color{red} \bar Z} & Z \end{smallmatrix} \right)$, 
 $\left( \begin{smallmatrix} \bar Z & \bar Z & {\color{red} Z}\\Z & Z &  {\color{red} \bar Z} \end{smallmatrix} \right)$ and 
 $\left( \begin{smallmatrix} \bar Z & {\color{red} Z} &  \bar Z\\ {\color{red} \bar Z} &  Z & Z \end{smallmatrix} \right)$. 
 The coefficient for each diagram is $1$, and their phases are, respectively $e^{ip}$, $1$ and  $e^{-ip}$.
		 
\item Finally, the interaction $\Tr( Z^3 \bar Z^3)$ gives two relevant diagrams $\left( \begin{smallmatrix} {\color{red} Z} 
& \bar Z & \bar Z\\ Z & Z & {\color{red} \bar Z} \end{smallmatrix} \right)$ and
$\left( \begin{smallmatrix} \bar Z & \bar Z & {\color{red} Z}\\{\color{red} \bar Z} & Z & Z \end{smallmatrix} \right)$. 
The coefficient for this diagram is $-1$, total of $-2$, and their phases are, respectively $e^{2ip}$ and $e^{-2ip}$.		 

\end{itemize}

The total factor for these diagrams is then
\be 
\begin{split}
\mathtt{ total\ factor}= f_{\bar Z}(p)=-4 + 2 (1 + e^{ip}+ e^{-ip}) - (e^{2ip}+ e^{-2ip}) & =8\sin^2 \frac{p}{2}\left(1-2 \sin^2 \frac{p}{2}\right) \; .\label{fbarz}
\end{split}
\ee

To find the Hamiltonian for these excitations, we note that, for $\phi_j$ representing the field at site $j$, the 5 diagrams give the following terms
(where the sum over $j$ is implicit)
\bea
&& -2\phi_j^2+2\phi_{j+1}\phi_j+2\phi_j \phi_{j-1}-\phi_{j-1}\phi_{j+1}-\phi_{j+1}\phi_{j-1}\cr
&=& -2\phi_j^2-4\phi_{j+1}\phi_j -2\phi_{j+1}\phi_{j-1}\cr
&=&2(\phi_{j+1}-\phi_j)^2-(\phi_{j+1}+\phi_{j-1}-2\phi_j)^2.
\eea
The discretized Hamitonian would be obtained by substituting $\phi_j=(b_j+b_j^\dagger)/\sqrt{2}$ and adding a free piece. But for matching we are 
in any case only concerned with small lattice momentum $p$. And note that the above interacting Hamiltonian is the discretization of 
\be
2(\phi')^2-(\phi'')^2\sim 2p^2\phi^2-p^4\phi^2. 
\ee
So at small $p$, only the first term, $2\phi'^2\rightarrow 2(\phi_{j+1}-\phi_j)^2$, remains, and this is, up to factor of $2$, 
the same Hamiltonian as for ${\cal N}=4$ SYM (therefore the same as in the case of the commutator insertions).

\subsubsection{Insertion of $\bar \phi^m$.}

We now need to do a similar calculation for the $\bar \phi^m$ insertion. We first expand the interaction Hamiltonian, the terms with two $\phi$'s only
(the only ones relevant in the dilute gas approximation) and obtain 
\bea
&&\Tr\left(\bar \phi^m \phi^m(2\bar Z^2 Z^2 -2Z^2 \bar Z^2-5\bar Z Z \bar Z Z +3 Z \bar Z Z \bar Z +Z \bar Z^2 Z+\bar Z Z^2 \bar Z)\right.\cr
&&\left.+\phi^m \bar \phi^m(2 Z^2\bar Z^2-2\bar Z^2 Z^2-5 Z \bar Z Z \bar Z +3 \bar Z Z \bar Z Z+Z\bar Z^2 Z +\bar Z Z^2 \bar Z)\right.\cr
&&\left. +2 \bar Z Z \bar \phi^m \bar Z Z \phi^m +2  Z \bar Z \bar \phi^m Z \bar Z \phi^m-2 \bar Z Z \bar\phi^m Z \bar Z \phi^m -2 Z \bar Z \bar\phi^m \bar Z
Z \phi^m\right)\;,
\eea
and we can rewrite this expression as
	\begin{align}
	\Tr & \left[\ \bar{\phi}^m\phi^m\left( 2 \bar Z^2 Z^2  - 5 \bar Z Z \bar Z Z + 5 Z \bar Z Z\bar Z + Z \bar Z \bar Z Z 
	+  \bar Z Z Z \bar Z \right) \phantom{\frac{1}{2}}\right.\nn \\
	& + \phi^m \bar{\phi}^m\left( 2 Z^2 \bar Z^2  - 5 Z \bar Z Z \bar Z + 5 \bar Z Z \bar Z Z + \bar Z Z Z \bar Z +  Z \bar Z \bar Z Z \right)\\
	&\left. - 6\bar \phi^m \phi^m Z^2 \bar Z^2 - 2 Z \bar \phi^m  Z \bar Z \phi^m \bar Z\phantom{\frac{1}{2}} \right]\;,\nn	
	\end{align}
where we have used the fact that $[\phi^m, Z]=0$ and $[\bar \phi^m, \bar Z]=0$, so  in particular 
$\Tr [Z \bar Z \bar \phi^m \bar Z Z \phi^m ] = \Tr [\phi^m \bar \phi^m \bar Z^2 Z^2 ] = \Tr [ \bar\phi^m \phi^m Z^2 \bar Z^2 ]$.

We write separately the contributions of each term to the spin chain Hamiltonian, obtaining 9 diagrams:
	\begin{itemize}
	\item[\bf \texttt{1st line}:]
	\item The terms $\Tr (\bar \phi \phi \bar Z Z \bar Z Z)$ and $\Tr (\bar \phi\phi Z\bar  Z Z \bar Z)$ don't contribute.
	\item The term $\Tr(\bar{\phi}\phi\bar Z^2 Z^2)$ gives the (relevant) diagram $\left( \begin{smallmatrix} 
	{\color{red} \phi } & \bar Z & \bar Z\\{\color{red} \bar \phi} & Z & Z \end{smallmatrix} \right)$ with coefficient $+2$ and trivial phase.
	\item The term $\Tr(\bar \phi \phi\bar Z Z Z \bar Z)$ gives the diagram $\left( \begin{smallmatrix}\bar Z &  
	{\color{red} \phi } & \bar Z\\{\color{red}\bar \phi} & Z & Z \end{smallmatrix} \right)$ with coefficient $+1$ and phase $e^{-ip}$
	\item The term $\Tr(\bar \phi \phi Z \bar Z \bar Z Z)$ gives the diagram	 $\left( \begin{smallmatrix}{\color{red} \phi}
	& \bar Z &  \bar Z\\Z & {\color{red}\bar \phi} & Z \end{smallmatrix} \right)$ with coefficient $+1$ and phase $e^{ip}$
	\item[\bf \texttt{2nd line}:]	
	\item The terms $\Tr (\phi \bar \phi Z \bar Z Z \bar Z)$ and $\Tr (\phi \bar \phi \bar Z Z\bar  Z Z)$ don't contribute
	\item The term $\Tr (\phi \bar \phi Z^2 \bar Z^2)$ gives the diagram $\left( \begin{smallmatrix} \bar Z & \bar Z 
	& {\color{red}  \phi }\\Z & Z & {\color{red} \bar \phi } \end{smallmatrix} \right)$ with coefficient $+2$ and trivial phase.
	\item The term $\Tr (\phi \bar \phi Z\bar Z^2 Z)$ gives the diagram $\left( \begin{smallmatrix} \bar Z & \bar Z 
	& {\color{red} \phi }\\Z &  {\color{red}\bar \phi} & Z  \end{smallmatrix} \right)$ with coefficient $+1$ and phase $e^{-ip}$
	\item The term $\Tr( \phi \bar \phi \bar Z  Z^2 \bar Z)$ gives the diagram	 $\left( \begin{smallmatrix} \bar Z 
	& {\color{red} \phi} &  \bar Z\\Z & Z & {\color{red}\bar \phi} \end{smallmatrix} \right)$ with coefficient $+1$ and phase $e^{ip}$	
	\item[\bf \texttt{3rd line}:]	
	\item The term $\Tr (\bar \phi \phi Z^2 \bar Z^2)$ gives the diagrams $\left( \begin{smallmatrix} \bar Z & \bar Z 
	& {\color{red}  \phi }\\{\color{red} \bar \phi } & Z &  Z\end{smallmatrix} \right)$ and $\left( \begin{smallmatrix} 
	{\color{red} \phi } & \bar Z & \bar Z \\Z & Z & {\color{red} \bar \phi } \end{smallmatrix} \right)$ with coefficient $-3$ 
	each and phases $e^{-2ip}$ and $e^{2ip}$, respectively.	
	\item The term $\Tr (Z \bar \phi Z \bar Z \phi \bar Z)$ gives the diagram $\left( \begin{smallmatrix} \bar Z & 
	{\color{red}  \phi }\ & \bar Z\\ Z & {\color{red} \bar \phi }\ &  Z\end{smallmatrix} \right)$ with coefficient $-2$ trivial phase	
	\end{itemize}
Therefore the total factor of the 9 diagrams above is 
\be 
\mathtt{ total\ factor} = f_{\bar \phi}(p)=2+2(e^{ip}+e^{-ip})-3(e^{2ip}+e^{-2ip})=
8\sin^2 \frac{p}{2}\left(5-6 \sin^2 \frac{p}{2}\right) \; .\label{fbarphi}
\ee

The Hamiltonian is found in the same way as for $\bar Z$ insertion. The 9 diagrams give the contributions (as before, the sum over $j$ is 
implicit)
\bea
&&2\phi_j^2+2\phi_j\phi_{j+1}+2\phi_j\phi_{j-1}-3\phi_{j+1}\phi_{j-1}-3\phi_{j-1}\phi_{j+1}\cr
&=& 10(\phi_{j+1}-\phi_j)^2-3(\phi_{j+1}+\phi_{j-1}-2\phi_j)^2\;,
\eea
which is the discretization for 
\be
10(\phi')^2-3(\phi'')^2\sim 10p^2\phi^2-3p^4\phi^4.
\ee
We see that, at small $p$, we have 5 times the result of the $\bar Z$ insertion.

\subsubsection{Insertion of $\phi^m$.}

We now look at the $\phi^m$ insertion. In this case, the insertion can propagate throughout the chain with no change. However, for the 
first nontrivial correction to the energy, planarity (which, if violated, carries a penalty of a $1/N^2$ factor) requires that the fields interact only 
``3 sites down with the corresponding 3 sites up", just like in the previous cases. 

We have now in principle the same 9 diagrams as in the case of the $\bar \phi^m$ insertion, just that now, because we can commute freely 
$\phi$ inside the $Z$'s, they all come from the same interaction. The vertices are all equivalent (under the commutation), and are equivalent to 
$\phi^m \bar\phi^m \bar Z^2 Z^2$. The result is that we have the same phase factors as for $\bar\phi$ insertion, 
but all have the same coefficient, which is now $-2/3$, namely:
\begin{itemize}
	\item  $\left( \begin{smallmatrix}  {\color{red} \bar  \phi } & \bar Z &\bar Z\\{\color{red}  \phi } & Z 
	&  Z\end{smallmatrix} \right)$, $\left( \begin{smallmatrix} \bar Z &  {\color{red} \bar \phi }  & \bar Z \\Z 
	&  {\color{red} \phi }  & Z \end{smallmatrix} \right)$ and $\left( \begin{smallmatrix} \bar Z & \bar Z &  
	{\color{red} \bar \phi }  \\Z & Z & {\color{red} \phi } \end{smallmatrix} \right)$ with trivial phase.
	\item  $\left( \begin{smallmatrix} \bar Z & \bar Z & {\color{red} \bar  \phi }\\ Z & {\color{red}  \phi } 
	&  Z\end{smallmatrix} \right)$ and $\left( \begin{smallmatrix} \bar Z & {\color{red} \bar \phi } 
	& \bar Z \\ {\color{red} \phi } & Z & Z\end{smallmatrix} \right)$ both with phases $e^{-ip}$.	
	\item  $\left( \begin{smallmatrix} \bar Z & {\color{red} \bar  \phi } & \bar Z\\ Z & Z &   {\color{red}  
	\phi }\end{smallmatrix} \right)$ and $\left( \begin{smallmatrix} {\color{red} \bar \phi }  & \bar Z & \bar Z \\ Z & {\color{red} \phi } 
	& Z\end{smallmatrix} \right)$ both with phases $e^{ip}$.		
	\item  $\left( \begin{smallmatrix} \bar Z & \bar Z & {\color{red} \bar  \phi }\\{\color{red}  \phi } 
	& Z &  Z\end{smallmatrix} \right)$ and $\left( \begin{smallmatrix} {\color{red} \bar \phi } 
	& \bar Z & \bar Z \\Z & Z & {\color{red} \phi } \end{smallmatrix} \right)$ with phases $e^{-2ip}$ and $e^{2ip}$, respectively.
	\end{itemize}	
Then the total factor of the 9 diagrams is
\be 
\mathtt{ total\ factor}=-\frac{2}{3}\left[3+2(e^{ip}+e^{-ip})+e^{2ip}+e^{-2ip}\right]=
\left[-6+16\sin^2 \frac{p}{2}\left(1-\frac{2}{3} \sin^2 \frac{p}{2}\right)\right] \; .
\ee
However, exactly like in the BMN case for ${\cal N}=4$ SYM \cite{Berenstein:2002jq}, 
there will be also other Feynman diagrams involving gauge fields and fermions, which 
come with a trivial phase (the scalars are not ``hopping" on the chain, but going straight up, and the gauge fields and fermions are 
making various connections to the scalar lines), and their result will be such that at $p=0$ we should have a total vanishing contribution, since 
the $p=0$ operator is a chiral primary field.  The result of this is that we can replace the coefficient of the trivial phase, $-(2/3)3$, with 
the one that cancels the rest at $p=0$, namely with $-(2/3)(-6)$, so that in the final expression for the total factor we remove the constant $-6$, and 
have 
\be 
\mathtt{ total\ factor}= f_\phi(p)=
\left[16\sin^2 \frac{p}{2}\left(1-\frac{2}{3} \sin^2 \frac{p}{2}\right)\right] \; .\label{fphi}
\ee

At small $p$, we get the same result as for the $\bar\phi^m$ insertion, except for an extra factor of 2/5.  

For the construction of the Hamiltonian, we do the same as before. We obtain from the diagrams above the contribution
\bea
&&-\frac{2}{3}[-6\phi_j^2+4\phi_{j+1}\phi_j+2\phi_{j+1}\phi_{j-1}]\cr
&=&4(\phi_{j+1}-\phi_j)^2-\frac{2}{3}(\phi_{j+1}+\phi_{j-1}-2\phi_j)^2\;,
\eea
which is the discretization of 
\be
4(\phi')^2-\frac{2}{3}(\phi'')^2\sim 4p^2 \phi^2-\frac{2}{3}p^4\phi^2.
\ee
Again at small $p$, we get the same Hamiltonian, except for an extra factor of 2/5. 

So in all these cases, except for overall factors, the interacting Hamiltonian at small lattice momentum $p$ is the same as in the ${\cal N}=4$ 
SYM case,
\be
[\phi_{j+1}-\phi_j]^2=[b_{j+1}+b^\dagger_{j+1}-b_j-b_j^\dagger]^2/2\;,
\ee
so the calculation of the energy gives the same square root, except for an overall factor of the interaction part.

\subsubsection{Eigenenergies and comparison with gravity dual}

The factors calculated above must still be multiplied by the same basic scalar diagram. 
In three dimensions the scalar propagator in position space is $P(x,y)= 1/(4\pi |x-y|)$.
The basic correction, a ``two-loop" Feynman diagram, which
we consider  has one operator ${\cal O}(0)$ at zero, and its conjugate $\bar{\cal O}(x)$ at $x$, is
integrated over a vertex at $y$, and has  3 propagators from the vertex to each of the operators, so $[P(y,0)]^3[P(x,y)]^3$. Since we 
want to compare with the free case (tree diagram), we divide the result by the tree result $[P(x,0)]^3$, obtaining \footnote{More generally, 
in a conformal theory we can 
use a conformal transformation to put the field theory on $\mathbb{R}\times\mathbb{S}^{d-1}$. Therefore, we calculate a general integral as
	\be 
	I(d,n)=\int \dd^d y\frac{1}{y^n (x-y)^n}= 	\int \dd y\frac{y^{d-1}}{y^n (x-y)^n}\int \dd \Omega_{d-1}\; ,
	\ee 
	and $\int \dd \Omega_{d-1}=2 \pi^{d/2}/\Gamma(d/2)$. For our case of $d=n=3$, we have
\bse
	\be 
	I= 4\pi \int \frac{\dd y}{y (x-y)^3}\; .
	\ee 
Doing the $y$ integral, we have an IR divergence and a divergence for $y=x$. We introduce the infrared cut-off $\Lambda$, 
and then  also $\Lambda=x-\texttt{max}(y)$, obtaining
\be 
I=\frac{8\pi}{|x|^3}\ln \Lambda\; .
\ee 	
\ese}
\bea
\frac{{\cal I}(x)}{{\cal I}^{\rm tree}(x)}&=&
\frac{4\pi^2}{k^2}\frac{|x|^3}{(4\pi)^3}\int d^3 y\frac{1}{y^3|x-y|^3}\cr
&=& \frac{4\pi^2}{k^2}\frac{8\pi}{(4\pi)^3} \ln |x|\Lambda+{\rm finite}
=\frac{1}{2k^2}\ln |x|\Lambda+{\rm finite}\;,
\eea
and we also have an $N^2$ factor coming from the sum over the 't Hooft double lines, for a total Feynman diagram factor, correcting the 
tree result, of (${\cal F}(x)\equiv \langle {\cal O}(x)\bar {\cal O}(0)\rangle$)
\be
\frac{{\cal F}(x)}{{\cal F}^{\rm tree}(x)}=1+f_i(p)\frac{N^2}{2k^2} \ln |x|\Lambda+{\rm finite}=1+f_i(p)\frac{\lambda^2}{2}\ln |x|\Lambda+{\rm finite}\;,
\ee
where $f_i(p)$ are the total factors $f_{\bar Z}(p), f_{\bar \phi}(p)$ and $f_\phi(p)$ calculated in the previous subsections. This is to be 
compared with the expected form
\be
\frac{{\cal F}(x)}{{\cal F}^{\rm tree}(x)}=(1+{\rm finite})\frac{|x|^{\Delta_{\rm tree}}}{|x|^{\Delta(\lambda)}}\simeq 1-\delta \Delta (\lambda)\ln |x|\Lambda
+{\rm finite},
\ee
to finally write 
\be
\Delta-J\simeq (\Delta-J)({\rm tree})+\delta \Delta(\lambda)=(\Delta-J)({\rm tree})-f_i(p)\frac{\lambda^2}{2}.
\ee
Note that the numerical coefficient multiplying $f_i(p)$ is different than in 3+1 dimensional ${\cal N}=4$ SYM, since the basic Feynman diagram 
is different. We also find the ``two-loop" $\lambda^2$ factor instead of $\lambda$, since now the vertex is sextic. 

{\bf Commutator insertions}. 

For completeness, we start by considering the case of $[\bar Z,Z]$ insertions, and then from (\ref{fbarzz}) we obtain
(since $p=2\pi n/J$ and $(\Delta-J)({\rm tree})=1$)
\be
\Delta -J \simeq 1-f_{[\bar Z,Z]}(p)\frac{\lambda^2}{2}= 1+ \frac{2N^2}{k^2}\sin^2\frac{p}{2}\simeq 1+2\pi^2\lambda^2\frac{n^2}{J^2}\;,
\ee
where the last result was for small $p$ (small $n$). 

Since the spin chain Hamiltonian at small $p$ is (up to a numerical factor) always the same 
as for 3+1 dimensional ${\cal N}=4$ SYM, the full result 
is the square root that has the above as the first correction, i.e. 
\be
\Delta-J=\sqrt{1-f_{[\bar Z,Z]}\lambda^2}=\sqrt{1+4\lambda^2\sin^2\frac{p}{2}}.
\ee


Like in the case of the  ABJM spin chain, all the more so since we have less than maximal supersymmetry, 
we expect the coefficient of $\sin^2 p/2$ to have some function of the 't Hooft coupling $\lambda=N/k$, i.e. we expect that
\be
\Delta-J=\sqrt{1+f_{[\bar Z,Z]}(\lambda)\sin^2\frac{p}{2}}\;,
\ee
such that for small $\lambda$, $f_{[\bar Z,Z]} (\lambda)\simeq 4 \lambda^2$.

For the case of {\em giant magnons}, $p$ is not infinitesimal anymore. In this case, we can neglect the 1 in the square root, and get
\be
\Delta - J \simeq \sqrt{f_{[\bar Z,Z]}(\lambda)}\left|\sin\frac{p}{2}\right|.
\ee
As described in  \cite{Araujo:2016jlx}, where giant magnons were analyzed in the GJV gravitational background,  
the dual giant magnon energy, given by the string action, 
is proportional to (since the Polyakov action is essentially the metric in string frame)  
\be
\frac{L^2_{\rm string}}{2\pi \a'}\propto \left(\frac{N}{k}\right)^{1/3}.
\ee
This means that at large $\lambda$, we have
\be
f_{[\bar Z,Z]}(\lambda)\propto \lambda^{2/3}.
\ee
Moreover, this result is valid for {\em any} type of spin chain insertion, corresponding to any direction for the giant magnon excitation. 

On the pp-wave, i.e. in the BMN limit, the scale factor is the same $L^2_{\rm string}$, so we should obtain in general at 
large $\lambda$
\be
f_i(\lambda)\propto \lambda^{2/3}.
\ee

{\bf Field insertions}

As we said, the physical case seems to be for insertions of the basic fields, rather than the commutators. For each case, we will have a 
set of different functions
\be
f_i(\lambda)=f_i(p)\lambda^2\;,
\ee
since we still have
\be
\Delta- J \simeq (\Delta-J)_0-f_i(p)\frac{\lambda^2}{2}\simeq (\Delta-J)_0-f_i \frac{\lambda^2}{2}\sin^2\frac{p}{2}\simeq (\Delta-J)_0
-f_i \frac{\lambda^2}{2}\frac{\pi^2n^2}{J^2}\;,
\ee
where the last two equalities are only valid for small $p$. 
But for matching with the pp-wave result, we must only consider small $p$, 
so we can neglect the $\sin^4 p/2$ terms in $f_i(p)$. 

The tree level value $(\Delta-J)_0$ for $\Delta-J$ is, as we already explained, 
1 for $D_a$ and $\bar Z$ insertions, and 1/2 for $\phi^m$ and $\bar\phi^m$ insertions. 
From matching with the pp-wave results (\ref{ea}) and (\ref{eb}) for $n=0$, we see that 
$D_a$ insertions correspond to $X^2,X^3,X^4$ insertions on the pp-wave, $\bar Z$ insertions to $X^1=u$ insertions, 
and $\phi^m, \bar\phi^m$ to $X^5,X^6,X^7,X^8$ insertions. 

Since, as we saw, $p^+=J/L^2_{\rm string}$, to translate from the pp-wave result we use the map, valid at $\lambda\rightarrow\infty$,
\be
\frac{n^2}{(\a' p^+)^2}=\left(\frac{L^2_{\rm string}}{\a'}\right)^2\frac{n^2}{J^2}\propto \lambda^{2/3}\frac{n^2}{J^2}.\label{fthmap}
\ee

\begin{itemize}

\item For $\bar Z$ insertions, from $f_{\bar Z}(p)$ in  (\ref{fbarz}), we get 
\be
\Delta-J\simeq 1-4\lambda^2\sin^2\frac{p}{2}\left(1-2\sin^2\frac{p}{2}\right)\;,
\ee
which from the spin chain Hamiltonian would be
\be
\sqrt{1-f_{\bar Z}(\lambda)\sin^2 p/2}\simeq 1-\frac{1}{2}f_{\bar Z}(\lambda)\sin^2\frac{p}{2}\simeq 1-\frac{\pi^2}{2}f_{\bar Z}(\lambda)\frac{n^2}{J^2}.
\ee
{\em at small $p$}, and then at small $\lambda$ we have $f_{\bar Z}(\lambda)\simeq 8\lambda^2$. By comparing with the small $p$ 
result on the pp-wave, eq. (\ref{ea}), we see that at large $\lambda$,
\be
f_{\bar Z}(\lambda)=-\frac{1}{\pi} \left(\frac{L^2_{\rm string}}{\a'}\right)^2\propto \lambda^{2/3}.
\ee

\item For $\bar \phi^m$ insertions, from $f_{\bar \phi}(p)$ in (\ref{fbarphi}), we get
\be
\Delta-J\simeq \frac{1}{2}-4\lambda^2\sin^2\frac{p}{2}\left(5-6\sin^2\frac{p}{2}\right)\;,
\ee
which from the spin chain Hamiltonian would be 
\be
\sqrt{\frac{1}{4}-\frac{1}{2}f_{\bar \phi}(\lambda)\sin^2 \frac{p}{2}}\simeq \frac{1}{2}-\frac{1}{2}f_{\bar \phi}(\lambda)\sin^2\frac{p}{2}
\ee
{\em at small $p$}, and at small $\lambda$ we have $f_{\bar \phi}(\lambda)\simeq 40\lambda^2$. By comparing with the 
small $p$ result on the pp-wave, eq. (\ref{eb}), we see that at large $\lambda$, 
\be
f_{\bar \phi}(\lambda)=-\frac{2}{\pi} \left(\frac{L^2_{\rm string}}{\a'}\right)^2\frac{n^2}{J^2}\propto \lambda^{2/3}.
\ee

\item For $\phi^m$ insertions, from $f_{\phi}(p)$ in (\ref{fphi}), we get
\be
\Delta-J\simeq \frac{1}{2}-8\lambda^2\sin^2\frac{p}{2}\left(1-\frac{2}{3}\sin^2\frac{p}{2}\right)\;,
\ee
which from the spin chain Hamiltonian would be 
\be
\sqrt{\frac{1}{4}-\frac{1}{2}f_{\phi}(\lambda)\sin^2 \frac{p}{2}}\simeq \frac{1}{2}-\frac{1}{2}f_{ \phi}(\lambda)\sin^2\frac{p}{2}
\ee
{\em at small $p$}, and at small $\lambda$ we have $f_{ \phi}(\lambda)\simeq 16\lambda^2$. By comparing with the 
small $p$ result on the pp-wave, eq. (\ref{eb}), we see that at large $\lambda$, 
\be
f_{ \phi}(\lambda)=-\frac{2}{\pi} \left(\frac{L^2_{\rm string}}{\a'}\right)^2\frac{n^2}{J^2}\propto \lambda^{2/3}.
\ee

\end{itemize}

Note that the functions $f_i(\lambda)$ are the same ones for the usual magnons (dual to the pp-wave) as they are 
for the giant magnons (corresponding to the calculation from \cite{Araujo:2016jlx}) since the scale in both gravity backgrounds is $L^2_{\rm string}$, 
which means we have the map (\ref{fthmap}).

In conclusion, we find matching with the pp-wave results, but only by introducing independent functions of coupling $f_i(\lambda)$, 
one for each field insertion.

\subsection{Sketch of field theory spin chain for open strings on D-branes}

We will now attempt to describe the spin chain for open strings on the giant graviton D4-branes, in the same way as it was done in 
\cite{Cardona:2014ora} for the ABJM case. We will only sketch the analysis, since we will see that we have an important puzzle, 
which we could not resolve.

From the point of view of the  analysis for closed strings excitation modes, we have the scalars $
Z,\bar Z$, charged under a $U(1)$ symmetry with charge $J$, and $
\phi^m, \bar\phi^m$, $m=1,2$, which we will split as $W, T$, forming an $SU(2)$ sector, and their conjugates, $\bar W,\bar T$. 

A maximal giant graviton D4-brane, wrapping the $\mathbb{CP}^2$ parametrized by 
$(\lambda,\theta,\phi,\sigma)$, thus with transverse coordinates $\rho$ (AdS radial coordinate) and angles 
$\a$ and $\psi$, where its motion is on the same direction $\psi$ as for the closed strings, 
will be described by a determinant operator made up of $Z$, the complex scalar charged under 
the rotation on $\psi$, i.e. 
\be
{\cal O}_{\rm g, max}=\epsilon_{m_1...m_N}\epsilon^{p_1...p_N}Z^{m_1}_{p_1}...Z^{m_N}_{p_N}.
\ee
A non-maximal giant graviton will correspond to a sub-determinant operator, and it is more complicated (Schur polynomials in general), so we will
stick with the maximum one. 

On the other hand, the open string must move in an {\em isometry direction} $A\in \mathbb{CP}^2$ parallel to the D4-brane giant, 
which restricts it to be 
$A=\sigma,\phi$ or a combination thereof (like $\sigma+\phi$). This corresponds indeed to the directions  in which 
we took the Penrose limits for the open string pp-wave.
That means that  the direction $A$ is charged under a different $U(1)$ charge $\bar J\neq J$. Since the direction $\a$, rescaled to $u$, 
corresponds (as we saw in the last section) to the $\bar Z$ insertion in field theory (it has the same energy vs. $\Delta-J$), it means that 
the rotation angle $A$ corresponds to (rotation of) a different 
$\phi^m, \bar \phi^m$ insertion. Let us define the complex scalar field charged under $\bar J$ 
to be $W$, and define it to have charge $\bar J=1/2$ (so that $\bar W$ 
has charge $\bar J=-1/2$). Then the vacuum of the open string will be  the open spin chain (with open matrix indices)
\be
[W^{\bar J}]^a_b.
\ee

Then the combined vacuum of the giant plus open string, of $\bar J$ units of open string  lightcone momentum $p^+$, is 
\be
|0,p^+\rangle ={\cal O}_{\rm g,max+open}=\epsilon_{m_1...m_N}\epsilon^{p_1...p_N}Z^{m_1}_{p_1}...Z^{m_{N-1}}_{p_{N-1}}[W^{\bar J}]^{m_N}_{p_N}.
\ee

Among the sites of the open string, $W^{\bar J}$, we must insert the excitations $\Psi^M$ of the open string, with their usual momentum factor, like 
\be
a_n^{\dagger M}|0,p^+\rangle \sim \sum_{l=0}^{J-1}e^{\frac{2\pi iln}{J}}
\epsilon_{m_1...m_N}\epsilon^{p_1...p_N}Z^{m_1}_{p_1}...Z^{m_{N-1}}_{p_{N-1}}[W^l\Psi^M W^{\bar J-l}]^{m_N}_{p_N}.
\ee
The problem is that the natural insertions $\Psi^M$ are: $D_a$, for $a=1,2,3$, with $\Delta-\bar J=1$, $\bar Z$ and $Z$, with $\Delta -\bar J=1/2$, 
$T$ and $\bar T$, with $\Delta-\bar J=1/2$, and $\bar W$, with $\Delta -\bar J=1$. 

Among the modes of the open strings, in the three Penrose limit directions analyzed, as we have explained in 
section \ref{openquant}, we have some modes that we cannot calculate
precisely, and five modes that are identical over the three different limits: the modes $X_i$, $i=1,2,3$, and the modes $Y_7,Y_8$, 
see (\ref{xiyi}) and (\ref{omegaopen}).  This means that no matter how we choose the field theory open spin chain, there should be 
a universality in the result.

But when looking at the modes of the open string on the pp-wave, we see that only the $D_a$ modes, corresponding to
the $X_i$ excitations, with energies $\omega_n^{(i)}$, 
match the $p=0$ values for the energy  ($\Delta-\bar J$ in the field theory),
 {\em and only if we rescale (reabsorb in the implicit $\mu$ scale for the energy) a $1/\sqrt{2}$ factor}
for $\omega$'s. 

After the rescaling by $\sqrt{2}$, the modes $Y_7,Y_8$ (with $E=\mu/\sqrt{6}$ at $n=0$) on the pp-wave have energies $E=\mu/\sqrt{3}$
at zero momentum, which however disagrees with both $\Delta-\bar J=1/2$ of $Z,\bar Z, T,\bar T$, or the $\Delta-J=1$ of $\bar W$. 
It is not clear how we could fix this mismatch.

\section{Conclusions}

In this paper we have studied Penrose limits of the GJV duality between the IR fixed point of ${\cal N}=2$ 
SYM-CS theories in 2+1 dimensions and a warped product geometry of the form $AdS_4 \times S^6$, where the six-sphere is squashed.  
On the gravitational side, we have calculated all the nontrivial Penrose limits along isometries of the background.

On the field theory side, after describing the Lagrangian and IR fixed point of the theory, 
we described a spin chain that corresponds to closed strings in one of the pp-waves, more concretely the Penrose 
limit along the R-symmetry direction. We obtained matching, only at the expense of introducing {\em independent}
functions $f_i(\lambda)$ for the various field theory insertions $\Phi_i$, and after restricting to small spin chain momentum $p$, so that 
$\sin^2 p/2\ll 1$. 

This situation is different from either of the more established examples, notably  3+1 dimensional ${\cal N}=4$ SYM (when there are no functions of the  coupling), or 2+1 dimensional ${\cal N}=6$ superconformal CS theory (ABJM model), where there is only one function. 
We attribute this feature to the fact that there is less supersymmetry (${\cal N}=2$ instead of the maximal ${\cal N}=8$) and more parameters, which makes the GJV duality more interesting. Even more so, since as we see, we can use a combination of the methods used in the ${\cal N}=4$ SYM and ABJM cases to analyze the model. It should be interesting to use methods based on quantum spectral curves to perform an analysis of these functions \cite{Gromov:2014eha, Cavaglia:2016ide}.

In the process of completing this work, an unresolved puzzle arose with respect to the open strings on D4-brane giant gravitons. 
A pp-wave corresponding to such a situation 
was found for each of the $\sigma,\phi, \sigma+\phi$ directions, but neither of these logical possibilities seem to match the result
obtained in the field theory at zero spin chain momentum $p$. We are unsure how to resolve this but, coupled with the fact that 
in the closed string case we have matching only for small $p$, suggests that this less supersymmetric case is more complicated 
and interesting than the standard ${\cal N}=4$ SYM and ABJM cases.

\section*{Acknowledgements} 

We would like to thank Carlos Cardona, Nakwoo Kim and 
Junchen Rong for discussions. TA was partially supported by PROPe-UNESP in the initial stages of the present work. GI 
was supported by FAPESP grant 2016/08972-0 and 2014/18634-9. The work of HN is supported in part by CNPq grant 304006/2016-5 and FAPESP grant 2014/18634-9. HN would also like to thank the 
ICTP-SAIFR for their support through FAPESP grant 2016/01343-7. GI and E\'{O}C would like to acknowledge the organizers of ``Recent Advances in T/U-dualities and Generalized Geometries" in Zagreb for providing a stimulating work environment, where the finishing touches to this paper were made.

\appendix

\section{3D superspace} 
\label{conventions}
\subsection{N=1 superspace}

In this section we review 3D $\mathcal{N}=1$ superspace. Our motivation to do so stems from the fact that recent treatments in the 
literature, for example \cite{Schwarz:2004yj, Gaiotto:2007qi}, have favoured working with components and the superspace 
conventions have not been comprehensive enough to stand alone. In this section, we will recapitulate some of the earlier
 work in this direction \cite{Gates:1983nr, Srivastava:1990cw}. 

We adopt the spacetime metric $\eta^{\mu \nu} = \textrm{diag} (-1, 1, 1)$, with gamma matrices, 
\be
(\gamma^0)^{\alpha}_{~\beta} = i \sigma_2, \quad (\gamma^1)^{\alpha}_{~\beta} = \sigma_1, \quad (\gamma^2)^{\alpha}_{~\beta} = \sigma_3, 
\ee
noting that $\gamma^{\mu} \gamma^{\nu} = \eta^{\mu \nu} + \epsilon^{\mu \nu \rho} \gamma_{\rho}$, with $\epsilon^{012} 
= \gamma^{012} = 1$. We next introduce $\epsilon^{\alpha \beta} = i \sigma_2, \epsilon_{\alpha \beta} = - i \sigma_2$, 
$\alpha, \beta = 1, 2$, allowing us to raise/lower indices on spinors and construct Lorentz invariants, 
\be
\bar{\psi}_{\alpha} = \epsilon_{\alpha \beta} \psi^{\beta}, \quad \psi^{\alpha} = \epsilon^{\alpha \beta} \bar{\psi}_{\beta} , 
\quad \bar{\psi} \chi = \bar{\psi}_{\alpha} \chi^{\alpha}. 
\ee 

To construct superfields, we introduce a two-component Majorana spinor comprising Grassmann coordinates $\theta^{\alpha}$, 
$\alpha = 1, 2$. With $\mathcal{N}=1$ supersymmetry, one can construct two different types of multiplets, namely scalar and gauge 
multiplets. We begin by defining a real scalar multiplet, consisting of two real scalars $\phi(x), C(x)$ and a Majorana spinor $\psi$: 
\be
\Phi (x, \theta) = \phi(x)+ i \bar{\theta} \psi(x) + \frac{i}{2} \bar{\theta} \theta C(x). 
\ee
We note that there are two real bosonic and two real fermionic degrees of freedom. The generator of the $\mathcal{N}=1$ supersymmetry 
transformation $Q^{\alpha}$ is given by, 
\be
i Q^{\alpha} = \frac{\partial}{\partial \bar{\theta}_{\alpha}} - i ( \gamma^{\mu} \theta)^{\alpha} \partial_{\mu}. 
\ee
The covariant superderivative is defined as 
\be
D^{\alpha} =  \frac{\partial}{\partial \bar{\theta}_{\alpha}} + i ( \gamma^{\mu} \theta)^{\alpha} \partial_{\mu}, \quad \bar{D}_{\alpha} 
= \epsilon_{\alpha \beta} D^{\beta}, \quad \textrm{such that} \quad 
\{ \bar{D}_{\alpha}, D^{\beta} \} = - 2 i (\gamma^{\mu})^{\beta}_{~\alpha} \partial_{\mu}. 
\ee
The supersymmetry variations follow from 
\be
\delta \Phi = \delta \phi + i \bar{\theta} \delta \psi  + \frac{i}{2} \bar{\theta} \theta \delta C = i  \bar{\epsilon}_{\alpha} Q^{\alpha} \Phi, 
\ee
which in terms of the component fields, leads to the variations \footnote{It is useful to recall the Fierz identity for Majorana spinors, 
which implies $ \bar{\epsilon} \gamma^{\mu} \theta \, \bar{\theta} \partial_{\mu} \psi = - \frac{1}{2} \bar{\theta} \theta \, \bar{\epsilon}
 \gamma^{\mu} \partial_{\mu} \psi$.}: 
\bea
\delta \phi = i \bar{\epsilon} \psi, \quad \delta \psi  = C \epsilon + \gamma^{\mu} \epsilon \partial_{\mu} \phi, \quad \delta C =
 i \bar{\epsilon} \gamma^{\mu} \partial_{\mu} \psi. 
\eea

From the scalar superfield $\Phi$ one can construct a scale invariant action of the form \footnote{We use conventions 
such that $\int \dd^2 \theta \bar{\theta} \theta = -1$.}, 
\be
S = \frac{1}{2} \int \dd^3 x \, \dd^2 \theta \bar{D} \Phi D \Phi = \int \dd^3 x \left(\frac{1}{2} \partial_{\mu} \phi \partial^{\mu} 
\phi - \frac{i}{2} \bar{\psi} \gamma^{\mu} \partial_{\mu} \psi + \frac{1}{2} C^2 \right).
\ee
To see that this is scale invariant, we assign dimensions, 
\be
[\Phi] = \frac{1}{2} ~~ \Rightarrow ~~[\phi] = - [\theta] =\frac{1}{2}, ~~ [\psi] = 1, ~~ [C] = \frac{3}{2}.  
\ee

Now that we have introduced the scalar multiplet, we can introduce the gauge multiplet. The gauge multiplet is contained in a Majorana spinor 
superfield $\Gamma^{\alpha}$, which consists of two two-component Majorana spinors, $\chi^{\alpha}$, $\lambda^{\alpha}$, a real scalar 
$a(x)$ and a vector potential $A_{\mu}(x)$, 
\be
\Gamma^{\alpha} ( x, \theta) = \chi^{\alpha} (x)  + \bar{\theta}_{\beta} [ \frac{1}{2} \epsilon^{\beta \alpha} a (x) 
+ (\gamma^{\mu})^{\beta \alpha} A_{\mu}(x) ] + i \bar{\theta} \theta \eta^{\alpha},  
\ee
where $\eta^{\alpha} = \lambda^{\alpha} - \frac{1}{2} ( \gamma^{\mu} \partial_{\mu} \chi)^{\alpha}$. Once again, we note that one has an equal 
number of bosonic and fermionic degrees of freedom, i. e. four real degrees. The infinitesimal gauge transformation of the spinor superfield 
is $\delta \Gamma^{\alpha} = - i D^{\alpha} \Phi$, where $\Phi$ is a real superfield. In terms of components, one finds $ \delta A_{\mu} 
= - \partial_{\mu} \phi, \delta \lambda = 0, \delta \chi = \psi, \delta a = - 2 C$. Noting that $ \lambda^{\alpha} = \frac{i}{2} i \bar{D}_{\beta} 
D^{\alpha} \Gamma^{\beta} |_{\theta = 0}$ is gauge invariant, the natural field strength superfield is, 
\be
W^{\alpha} = \frac{i}{2} \bar{D}_{\beta} D^{\alpha} \Gamma^{\beta}, 
\ee
where gauge invariance follows from the identity $\bar{D}_{\beta} D^{\alpha} D^{\beta} = 0$. The Chern-Simons action is obtained from the action 
\be
S_{CS} = \int \dd^3 x \dd^2 \theta  \bar{\Gamma} W = \int \dd^3 x \left( \epsilon^{\mu \nu \rho} A_{\mu} \partial_{\nu} A_{\rho} - \bar{\lambda} \lambda \right).  
\ee
where we have evaluated the action in the supersymmetric gauge $\bar{D} \Gamma = 0$, which corresponds to setting 
$ a= 0, \partial^{\mu} A_{\mu} = 0$ and $\lambda = \gamma^{\mu} \partial_{\mu} \chi$. The generalisation to the non-Abelian case is straightforward. 

\subsection{N=2 Chern-Simons-matter theories}

In this section, we follow the conventions of \cite{Benna:2008zy}. The three dimensional spinor group for a Minkowski spacetime 
with metric $\eta_{\m\n}=\texttt{diag}(-++)$ is $Spin(1,2)\simeq Sl(1,2)\simeq SU(1,1)$.

Field theories with ${\cal N}=1$ supersymmetries are formulated by a two real components Majorana spinor, but this amount of supercharges is 
not enough to provide holomorphy properties to these theories. With ${\cal N}=2$ supersymmetries, $4$ real components, the situation is easier, 
and it is the case we consider here \cite{Aharony:1997bx}. The Dirac matrices are $(\gamma^\mu)_a^{\phantom{a}b}=(i\s^2, \s^1,\s^3)$, that is
\be 
\gamma^0=
\begin{pmatrix}
0 & 1\\
-1 & 0
\end{pmatrix}\; , \quad
\gamma^1=
\begin{pmatrix}
0 & 1\\
1 & 0
\end{pmatrix}\; , \quad
\gamma^2=
\begin{pmatrix}
1 & 0\\
0 & -1
\end{pmatrix}\; .
\ee
and satisfy $\gamma^\m \gamma^\n = \eta^{\m\n} + \epsilon^{\m\n\r}\gamma_\r$ and $\{\g_\m,\g_\n \}=2\eta_{\m\n}$.

A Majorana spinor transforms in the fundamental representation of $Sl(1,2)$, therefore, for the ${\cal N}=2$ case, we combine two 
Majorana spinors into a Dirac spinor that transforms in the fundamental representation of $SU(1,1)$ \cite{Gates:1983nr}. A generic 
Dirac spinor can be written as $\psi=(\psi^1\ \psi^2)^T\in \mathbb{C}^2$, and its dual is defined as $\psi^a:=\epsilon^{a b}\psi_b\in 
\mathbb{C}^2$, where $\epsilon^{12}=-\epsilon^{21}=1$, that is $\epsilon^{\a\b}=i\sigma^2$. Moreover, we define $\epsilon_{ab}=
-i \s^2$, that satisfies $\epsilon_{a c}\epsilon^{c b}=\delta_a^b$, or more generally 
\be 
\epsilon^{ab}\epsilon_{cd}=\delta_d^a \delta^b_c - \delta_c^a \delta^b_d\; , \quad  \epsilon^{ab}\epsilon^{cd}=\delta^{ac} \delta^{bd} 
- \delta^{ad} \delta^{bc} \; , \quad  \epsilon_{ab}\epsilon_{cd}=\delta_{ac} \delta_{bd} - \delta_{ad} \delta_{bc}\; .
\ee
In fact, it easy to see that the gamma matrices $(\gamma^\mu)_{ab}=\epsilon_{a c}(\gamma^\mu)_b^{\phantom{a}c}=
(-\mathbb{1}, -\s^3, \s^1)$ are symmetric. 

The action of the Lorentz group on these spinors is
\be 
\begin{split}
\psi_a &\ \ \to\ \ {\cal M}_a^{\phantom{a}b}\psi_b\\
\psi^a &\ \ \to\ \ \psi^b ({\cal M}_a^{\phantom{a}b})^{-1}\equiv \psi^b {\cal M}_b^{\phantom{b}a}\; ,
\end{split}
\ee
and with these definitions, the product $\psi\xi:=\psi^a\xi_a=\xi\psi$ is Lorentz invariant. Furthemore, the Dirac and
 Majorana conjugates are defined, respectively by 
\be 
\begin{split}
\bar{\psi}^a & :=(\psi^\dagger)^b(\gamma_0)_b^{\phantom{b}a}\\
(\psi^C)^a & := (\psi^T)^b{\cal C}_b^{\phantom{b}a}\; .
\end{split}
\ee
where the charge conjugation matrix is ${\cal C}:=-i\g_0=\s^2$. The Dirac spinors can be decomposed as
\be 
\psi_a=\psi^1_a+i \psi_a^2\ \; ,
\ee
where $\psi^i$, $i=1,2$ are two Majorana spinors, and also $\bar{\psi}_a=\epsilon_{ab}\bar{\psi}^b$. Also, we define the
 contractions $\psi\g_\m\bar{\theta}:= \psi^a (\gamma_\m)_{a b}\bar{\theta}^b$.

It is easy to show that following identities
\begin{align}
&\theta_a\theta_b= \frac{1}{2}\epsilon_{ab}\theta^2\; , \quad \theta^a\theta^b=- \frac{1}{2}\epsilon^{ab}\theta^2\; , \quad 
(\theta \bar{\theta})^2=-\frac{1}{2}\theta^2\bar{\theta}^2\; ,\nn\\
&\theta\g^\mu \bar{\theta} \theta\g^\n \bar{\theta}=\frac{1}{2}\eta^{\m\n}\theta^2\bar{\theta}^2\; ,
\quad (\theta \bar{\theta}) \theta \g^\m \bar{\theta}=0\; , \quad (\theta \bar{\theta})^\dagger=-\theta \bar{\theta}\; , \label{iden.}\\
&(\bar{\theta} \bar{\lambda})^\dagger= (\theta^\dagger \lambda^\dagger)^\dagger=-\theta \lambda\; , \quad (\bar{\theta} 
\bar{\theta})^\dagger= (\theta^\dagger \theta^\dagger)^\dagger=-\theta \theta \; ,\nn\\
&(\theta \g^\m \bar{\psi})^\dagger= \psi \g^\m \bar{\theta}\; , \quad \quad (\theta \g^\m \bar{\theta})^\dagger= \theta \g^\m \bar{\theta}\;
\nn\\
& \theta\lambda (\theta\gamma^\m \bar \theta)=-\frac{1}{2}\theta^2 (\lambda\gamma^\m \bar{\theta})\; , 
 \quad (\theta \gamma^\m)_a(\gamma^\n \bar{\theta})^a = \theta\bar \theta \eta^{\m\n} - \epsilon^{\m\n\r}\theta \gamma_\r \bar \theta\; .
\end{align}

The susy generators and superderivatives in the superspace are given by
\bse
\be 
\begin{split}
Q_a & = \d_a-i(\g^\m \bar{\theta})_a\d_\m \; , \qquad \bar{Q}_a =- \bar{\d}_a+i(\theta\g^\m)_a\d_\m\\
D_a & = \d_a+i(\g^\m \bar{\theta})_a\d_\m \; , \qquad \bar{D}_a =- \bar{\d}_a-i(\theta\g^\m)_a\d_\m\; ,
\end{split}
\ee
and we can raise indices using $\d^a=-\epsilon^{ab}\d_b$. In the coordinates $y^\mu=x^\mu + i\theta \gamma^\mu \bar{\theta}$, we have
\be 
\begin{split}
Q_a & = \d_a \; , \qquad \bar{Q}_a =- \bar{\d}_a+2i(\theta\g^\m)_a\frac{\d}{\d y^\m}\\
D_a & = \d_a+2i(\g^\m \bar{\theta})_a\frac{\d}{\d y^\m} \; , \qquad \bar{D}_a =- \bar{\d}_a\; .
\end{split}
\ee
\ese
Under these definitions, it's easy to show that the chiral superfield $\Phi$, defined by $\bar{D}_a\Phi = 0$ has the most 
general expansion $\Phi(y)=\phi(y) + \sqrt{2}\theta \psi(y)+\theta\theta F(y)$, that is,
\bse
\be 
\Phi(x)=\phi+i(\theta \g^\m \bar{\theta})\d_\m \phi-\frac{1}{4}\theta^2 \bar{\theta}^2\square\phi + \sqrt{2}\theta \psi(x)
-\frac{i}{\sqrt{2}}\theta^2 \d_\m \psi \g^\m \bar{\theta}+\theta\theta F(x)\; .
\ee
In addition, we can define the anti-chiral superfield $\Phi^\dagger\equiv \bar{\Phi}$, such that $D_\a \bar{\Phi}=0$, 
then $\bar{\Phi}(\bar{y})=\bar{\phi}(\bar{y}) - \sqrt{2}\bar{\theta} \bar{\psi}(\bar{y})-\bar{\theta}\bar{\theta} \bar{F}(\bar{y})$, that is
\be 
\bar{\Phi}(x)=\bar{\phi}-i(\theta \g^\m \bar{\theta})\d_\m \bar{\phi}-\frac{1}{4}\theta^2 \bar{\theta}^2\square\bar{\phi} 
- \sqrt{2}\bar{\theta}\bar{\psi}(x)-\frac{i}{\sqrt{2}}\bar{\theta}^2  \theta \g^\m \d_\m \bar{\psi}-\bar{\theta}\bar{\theta} \bar{F}(x)\; .
\ee
\ese

The vector field $\cal V$ in the Wess-Zumino gauge is given by 
\bse
\be 
{\cal V}(x)= 2i \theta \bar{\theta} \s + 2 \theta \gamma^\m \bar{\theta} A_\m + i \sqrt{2}\theta^2 \bar{\theta}\bar{\chi} 
- i \sqrt{2}\bar{\theta}^2 \theta\chi + \theta^2 \bar{\theta}^2 D\; ,
\ee
and using the identities (\ref{iden.}), we can easily show that ${\cal V}={\cal V}^\dagger$ and 
\be
\begin{split} 
{\cal V}^2 & = 2(\s^2+ A^\m A_\m) \theta^2 \bar{\theta}^2\\
{\cal V}^3 & =0\; .
\end{split}
\ee
\ese

\paragraph{$\spadesuit$  Chern-Simons Lagrangian}

The superspace Lagrangian for the ${\cal N}=2$ nonabelian Chern-Simons is
\be 
{\cal L}_{CS}= \frac{k}{4\pi}\int\dd^4\theta \int_0^1\dd t \frac{i}{2}\text{Tr} {\cal V}\bar{D}^a W_a \; 
\ee
where
\be 
W_\a=e^{-t {\cal V}}D_a e^{t {\cal V}}= t D_a {\cal V} -t^2 {\cal V} D_a {\cal V}  + \frac{t^2}{2} D_a {\cal V}^2\; ,
\ee
therefore
\be 
{\cal L}_{CS}= \frac{k}{4\pi}\int\dd^4\theta \frac{i}{4}\Tr \left(  {\cal V}\bar D^a D_a {\cal V} - \frac{2}{3}{\cal V}\bar 
D^a( {\cal V} D_a {\cal V}) +\frac{1}{3}{\cal V}\bar D^a D_a {\cal V}^2 \right)\; .
\ee

Therefore, we have the following $D$-terms 
\bse
\begin{align}
\left.\phantom{\frac{1}{2}}\text{Tr} {\cal V}\bar{D}^a D_a {\cal V}\right|_{\theta^2 \bar{\theta}^2 } & = 4\text{Tr}(2 i D \s 
-  i \epsilon^{\m\n\r}A_\m \d_\n A_\r +  \chi \bar{\chi})\\
\left.\phantom{\frac{1}{2}}\text{Tr} {\cal V}\bar{D}^a( {\cal V} D_a {\cal V})\right|_{\theta^2 \bar{\theta}^2 }  & 
= 4\text{Tr}\left[ i \s (\s^2 + A_\m A^\m) + \epsilon^{\m\n\r}A_\m A_\n A_\r\right] \\
\left.\phantom{\frac{1}{2}}\text{Tr} {\cal V}\bar{D}^a D_a {\cal V}^2 \right|_{\theta^2 \bar{\theta}^2 } & =8i \text{Tr}[ \s (\s^2 + A_\m A^\m) ]\; .
\end{align}
\ese

Using these results, the ${\cal N}=2$ Chern-Simons action is
\be 
S_{CS} = \frac{k}{4\pi}\int \dd^3 x\text{Tr}\left[\left(\epsilon^{\m\n\r}A_\m \d_\n A_\r + \frac{2 i }{3}A_\m A_\n A_\r \right)+ i \bar{\chi}\chi -2 D\s \right]
\ee

\paragraph{$\clubsuit$ Charged Matter Lagrangian}

Let us consider, for the sake of generality, a family of chiral superfields, that is $\Phi^i$, $i=1, \cdots, n$.  The gauged Lagrangian for charged matter fields is
\be 
{\cal L}_m=-\int \dd^4 \theta\sum_i \bar{\Phi^i}e^{\cal V} \Phi^i \equiv- \left( \bar{\Phi}^i \Phi^i + \bar{\Phi}^i{\cal V} \Phi^i 
+ \frac{1}{2}\bar{\Phi}^i{\cal V}^2 \Phi^i \right)_{\theta^2 \bar{\theta}^2}\; .
\ee
Therefore
\bse
\be
\Phi^{i \dagger} {\cal V}^2\Phi^i = 2\phi^{i\dagger}\left(\s^2+A^\m A_\m \right)\phi^i\ \theta^2 \bar{\theta}^2
\ee
and also
\begin{align}
\left.\Phi^{i\dagger} {\cal V} \Phi^i \right|_{\theta^2\bar{\theta}^2} &  = i\left(\phi^{i\dagger} A_\m \d^\m \phi^i 
- \d_\m \phi^{i\dagger} A^\m \phi^i \right)+i \phi^{i\dagger} \chi \psi^i + i \bar{\psi}^i \bar{\chi} \phi^i-i\bar{\psi}^i \s \psi^i\\
& \phantom{XX} +\bar{\psi}^i \g^\m A_\m \psi^i +\phi^{i\dagger} D\phi^i\nn \\
\left.\Phi^{i\dagger} \Phi^i \right|_{\theta^2\bar{\theta}^2} &  =\d_\m \phi^{i\dagger} \d^\m \phi^i 
+i\bar{\psi}^i \g^\m \d_\m\psi^i -\bar{F}^i F^i+ \texttt{(total derivative)}\; .
\end{align}
\ese
Then
\be 
\begin{split}
{\cal L}_m  = -\Tr & \left[ (D_\m \phi)^{i\dagger} D^\m \phi^i + i\bar{\psi}^i \gamma^\m D_\m \psi^i -\bar{F}^iF^i 
+ \phi^{i\dagger} D \phi^i + \phi^{i\dagger} \s^2 \phi^i -i\bar{\psi}^i \s \psi^i\right.\\
& \left.+ i \phi^{i\dagger} \chi \psi^i + i \bar{\psi}^i \bar{\chi}\phi^i \right]
\end{split}
\ee
where the convariant derivative is $D_\m=\d_\m-i[A_\m, \cdot]$.

\paragraph{$\maltese$ Superpotential:}
The superpotential Lagrangian is
\be 
\begin{split}
{\cal L}_{sp} & =-\int \dd^2 \theta {\cal W}(\Phi) - \int \dd^2 \bar{\theta} \overline{{\cal W}(\Phi)} \\
&  =- \Tr \left( \frac{\d {\cal W}(\phi)}{\d \phi^i} F^i +  \frac{\d \overline{{\cal W}(\phi)}}{\d \bar{\phi}^i} \bar{F}^i 
-   \frac{1}{2}\frac{\d^2 {\cal W}(\phi)}{\d \phi^i \d \phi^j}\psi^i \psi^j - \frac{1}{2}\frac{\d^2 \overline{ {\cal W}(\phi)}}{\d \bar{\phi}^i 
\d \bar{\phi}^j}\bar{\psi}^i \bar{\psi}^j \right)\; .
\end{split}
\ee

Consider that the matter fields $X$ are Lie-algebra valued. Therefore, the fields in the gauge multiplet, denoted collectively by $G$, 
act in the matter fields $X$ adjointly, for example $D \phi = [D,\phi]=(f^{abc} D^a \phi^b )T^c$ (in this particular case, let the Latin 
indices denote indices in the algebra). Moreover, we have
\be 
\begin{split}
\text{Tr}(\bar{X}G X) &= \text{Tr}(\bar{X}[G, X]) =  \frac{1}{2}f^{\a\b\g}\bar{X}^a G^b X^c\\
\text{Tr}(\bar{X}G^2 X) &= \text{Tr}(\bar{X}[G,[G, X]]) =  \frac{1}{2}f^{eab} f^{ecd}\bar{X}^a G^b G^c X^d\; ,
\end{split}
\ee
where the trace is normalized as $2 \text{Tr}(T^a T^b)=\delta^{ab}$. Integrating out the auxiliary fields, we find 
\be
\begin{split}
\s^a & = - \frac{4\pi}{k}\text{Tr}\left( \phi^{i\dagger} T^a \phi^i \right)\; , \qquad F^i =-\frac{\d {\cal \bar{W}}(\phi^\dagger)}
{\d \phi^{i\dagger}}\; , \qquad F^{i\dagger}=-\frac{\d {\cal W}(\phi)}{\d \phi^i }\\
\chi^a & = \frac{8\pi}{k}\text{Tr} \left( \bar{\psi}^i T^a \phi^i \right)\; , \qquad
\bar{\chi}^a =  \frac{8\pi}{k}\text{Tr} \left( \phi^{i\dagger} T^a \psi^i \right)\; ,
\end{split}
\ee
so that $k\text{Tr}(D\s)=-2\pi \text{Tr}(\phi^{i\dagger} D \phi^i) $.

Therefore, the ${\cal N}=2$ Chern-Simons-matter Lagrangian is
\begin{align}
{\cal L}  = & \text{Tr}\left[ \frac{k}{4\pi}\epsilon^{\m\n\r} \left(A_\m \d_\n A_\n + \frac{2 i}{3}A_\m A_\n A_\r \right) + \frac{k}{4\pi}(i \bar{\chi}\chi -2D \sigma)
-\left( D_\m \phi^{i\dagger} D^\m \phi^i + i\bar{\psi}^i \gamma^\m D_\m \psi^i \right)\right]  \nn\\
&  - \frac{8\pi i}{k}\text{Tr}\left(\bar{\psi}^i T^a \phi^i \right) \text{Tr}\left(\phi^{j\dagger} T^a \psi^j \right) 
- \frac{16 \pi^2}{k^2}\text{Tr}\left( \phi^{i\dagger} T^a \phi^i \right)  \text{Tr} \left( \phi^{j\dagger} T^b \phi^j \right)
 \text{Tr} \left( \phi^{k\dagger} T^a T^b \phi^k \right)\nn\\
&  - \frac{4\pi i}{k}\text{Tr}\left(\phi^{i\dagger} T^a \phi^i \right)\text{Tr} \left(\bar{\psi}^j T^a \psi^j \right)\nn\\
&+ \Tr \left( \frac{\d {\cal W}(\phi)}{\d \phi^i} \frac{\d \overline{{\cal W}(\phi)}}{\d \bar{\phi}^i} 
+   \frac{1}{2}\frac{\d^2 {\cal W}(\phi)}{\d \phi^i \d \phi^j}\psi^i \psi^j+\frac{1}{2}\frac{\d^2 \overline{ {\cal W}(\phi)}}{\d \bar{\phi}^i 
\d \bar{\phi}^j}\bar{\psi}^i \bar{\psi}^j \right)\; .\label{csm-lag}
\end{align}

\bibliographystyle{utphys}
\bibliography{GJVPenroserefs}{}

 
\end{document}